\pdfoutput=1

\documentclass[a4paper,11pt]{article}
\usepackage{jcappub} 
\usepackage{aas_macros}
 \usepackage{multirow} 
 \usepackage{lineno}

\title{Sensitivity of the Cherenkov Telescope Array to the detection of axion-like particles at high gamma-ray opacities}
\author[a]{Manuel Meyer}
\author[a]{Jan Conrad}

\affiliation[a]{The Oskar Klein Center for CosmoParticle Physics, Department of Physics, Stockholm University, Albanova, SE-10691 Stockholm, Sweden}
\emailAdd{manuel.meyer@fysik.su.se}
\emailAdd{conrad@fysik.su.se}

\arxivnumber{1406.5972}

\abstract{
Extensions of the Standard Model of particles commonly predict the existence of axion(-like) particles (ALPs)
that could be detected through their coupling to photons in external magnetic fields. 
This coupling could lead to modifications of $\gamma$-ray spectra from extragalactic sources. 
Above a certain energy, the $\gamma$-ray flux should be exponentially damped
due to the interaction with photons of background radiations fields. 
ALPs, on the other hand, propagate unimpeded over cosmological distances 
and a reconversion into $\gamma$-rays 
could lead to an additional component in the spectra.

Here, we present the sensitivity of the proposed Cherenkov Telescope Array (CTA) to detect this spectral hardening.
Using the full instrumental response functions of CTA, a 
combined likelihood analysis of four $\gamma$-ray sources
shows that a significant detection of the ALP signal is possible 
for couplings $g_{a\gamma} \gtrsim 2\times10^{-11}\,\mathrm{GeV}^{-1}$
and ALP masses $m_a \lesssim 100\,\mathrm{neV}$. 
We discuss the dependency of these values on different model assumptions 
and magnetic-field scenarios and identify the best observation strategy
 to search for an ALP induced boost of the $\gamma$-ray flux.
}

\begin{document}

\maketitle

\flushbottom

\section{Introduction} 
\label{sec:intro}
Theories beyond the Standard Model of particle physics 
commonly predict the existence of pseudo-Nambu-Goldstone bosons (pNGB)
that might be very light and only very weakly coupled to Standard-Model particles \cite{jaeckel2010}.
For instance, such particles typically arise in 
string theories \cite[e.g.][]{witten1984,ringwald2014}
and could constitute all of the dark matter content through their non-thermal 
production in the early Universe \cite[e.g.][]{preskill1983,marsh2011,arias2012}.
One prominent example is the axion that explains the non-observation of the electric dipole moment of the neutron, i.e. the strong CP problem in QCD
\cite{pq1977,weinberg1978,wilczek1978}. 
These spin-0 
bosons could be detected through their coupling to photons 
described by the Lagrangian \cite[e.g.][]{raffelt1988}
\begin{equation}
\mathcal{L}_{a\gamma} = -\frac 1 4 g_{a\gamma}F_{\mu\nu}\tilde F^{\mu\nu}a = g_{a\gamma}\, \mathbf{E}\,\mathbf{B}\, a,
\label{eq:lagr-alp}
\end{equation}
where $g_{a\gamma}$ is the coupling constant (with dimension $\mathrm{energy}^{-1}$),
$F_{\mu\nu}$ is the electromagnetic field tensor, $\tilde F^{\mu\nu} = 1/2 \epsilon^{\mu\nu\rho\sigma}F_{\rho\sigma}$ 
is its dual, and $a$ denotes the field strength of the pNGB.
As the right-hand side of the above equation shows, pNGBs couple to the electric field of the photon $\mathbf E$ in the presence 
of an external magnetic field $\mathbf B$. 
The photon-axion coupling is directly proportional to its mass $m_a$ since both quantities are related to the breaking scale $f_a$ of an additional gauge symmetry 
(see the review in ref. \cite{peccei2008}).
In more general theories, the mass and coupling are unrelated and one speaks of axion-like particles (ALPs)
on which we focus in the following. 

Owing to the ubiquitous presence of magnetic fields along the line of sight 
to blazars, active galactic nuclei (AGN) with their jet closely aligned to the line of sight,
photon-ALP oscillations could leave an imprint on $\gamma$-ray spectra of these sources.
The conversions would lead to different observable spectral features:

(i) around a critical energy (see section \ref{sec:alp}) the photon-ALP mixing 
becomes maximal and independent of energy. This leads to a drop in the $\gamma$-ray flux
which is accompanied by oscillatory features in the spectrum that depend on the structure of the ambient magnetic field. 
The search for such irregularities in $\gamma$-ray spectra has been used to place limits 
on the photon-ALP coupling \cite{hess2013:alps}.
(ii) Photon-ALP oscillations could lead to a reduced optical depth, 
$\tau$, of the Universe.
The $\gamma$-ray flux of extragalactic sources is exponentially attenuated by the factor $\exp(-\tau)$
due to pair production of $\gamma$ rays with photons of background radiation fields, $\gamma\gamma\to e^+e^-$.
The optical depth increases monotonically with 
the redshift $z$ of the AGN, the $\gamma$-ray energy $E$,
and the photon density of the background radiation field  (see the recent review of ref. \cite{dwek2013}).
The cross section for pair production peaks around a wavelength $\lambda_\ast \sim 1.24 (E / \mathrm{TeV})\,\mu$m \cite[e.g.][]{guy2000},
and, consequently, the most important background radiation field for the interaction 
with very high energy (VHE, $E \gtrsim 100\,\mathrm{GeV}$) $\gamma$-rays is the extragalactic background light (EBL)
which stretches from ultraviolet (UV) to far infrared wavelengths. 
The EBL comprises the emitted starlight and starlight absorbed and re-emitted by dust in galaxies
integrated over the history of the Universe \cite[e.g.][]{hauser2001}.
The exact level of its photon density remains unknown since direct measurements suffer from the contamination 
of foreground emission \cite{hauser1998}. 
However, firm lower limits can be derived from integrated galaxy number counts \cite{madau2000,fazio2004} and 
recent models of the EBL predict photon densities close to these lower limits \cite[e.g.][]{franceschini2008,kneiske2010,dominguez2011,gilmore2012}.
Nevertheless, evidence exists that even these models predict a too large photon density:
Not only is the number of observations of blazar in the optical thick regime (i.e., high values of $\tau$)
increasing (e.g., refs. \cite{1es1101h2356hess2006,3c279magic2008,pg1553hess2008,pks1424veritas2010}), 
also the
observed spectral indices\footnote{
Observed $\gamma$-ray spectra
are often satisfactorily described with power laws, $\phi(E) \propto E^{-\Gamma}$, where $\Gamma$ is the spectral index.}
do not seem to follow the expected softening with redshift as predicted by EBL absorption
 \cite{deangelis2009,deangelis2011}.
The absorption corrected VHE spectra of several blazars furthermore show a spectral flattening at the highest energies \cite{dominguez2011alps}. 
A statistical analysis of 50
blazar spectra of which 7 have data points that correspond to $\tau > 2$
shows a $~4\,\sigma$ indication for a too strong attenuation in the optical thick regime \cite{horns2012},
even though an EBL model is applied that is designed to predict a minimal absorption 
at TeV energies \cite[henceforth KD2010]{kneiske2010}.
Additionally, evidence is found for a redshift dependent spectral hardening at $\tau = 1$ 
in a sample of $\gamma$-ray spectra which can also be interpreted as an over-estimation of the EBL photon density \cite{rubtsov2014}.
Photon-ALP oscillations could be an explanation for these indications,
since ALPs propagate unimpeded over cosmological distances.
ALPs re-converting into $\gamma$ rays could lead to a boost in the observed photon flux \cite[e.g.][]{deangelis2007,mirizzi2007,sanchezconde2009,deangelis2011}. 

Here, we will investigate the sensitivity to detect this boost with the future Cherenkov Telescope Array (CTA). 
CTA is expected to have a sensitivity  a factor of 10 or more better than 
currently operating imaging air Cherenkov 
telescopes (IACTs), H.E.S.S. \cite{aharonian2006crab}, MAGIC \cite{magic2012crab}, and VERITAS \cite{holder2008veritas}, 
in the energy range between tens of GeV up to $\sim 100$ TeV with an energy resolution
of about 10 - 15\,\% and arcmin scale angular resolution \cite{cta2011}.
The broad energy coverage 
will enable the simultaneous 
observation of blazars in the optical thin ($\tau <  1$) and optical thick regime ($\tau \gg 1$), essential
to detect an enhanced $\gamma$-ray flux at highest energies.
Observations will be simulated with and without an ALP contribution 
making use of the full instrumental response functions (IRFs) which are derived from 
dedicated simulations \cite{bernloehr2013}.
To quantify the sensitivity, a likelihood ratio test will be used as described in ref. \cite{meyer2014}.

The article is structured as follows. In section \ref{sec:alp} we review the basic phenomenology for photon-ALP mixing
and the magnetic fields present along the line of sight. We will combine the results from a number of simulated blazar observations
and the considered source sample and the magnetic-field scenario for each source are described in section \ref{sec:sources}.
The simulations and the likelihood ratio test are introduced in section \ref{sec:method} before presenting our results in section \ref{sec:results}.
We discuss our model assumptions and source selection in section \ref{sec:disc} before concluding in section \ref{sec:concl}.
Throughout this article we will assume a standard $\Lambda$CDM cosmology with $h = 0.72$, $\Omega_m$ = 0.3 , and $\Omega_\Lambda = 0.7$ 
as chosen in the EBL modelling of ref. \cite{kneiske2010}\footnote{
The results presented here will only weakly depend on the chosen cosmology.
}.

\section{Photon-ALP mixing and astrophysical magnetic fields} 
\label{sec:alp}
The mixing between ALPs and photons that arises through the Lagrangian in eq. \eqref{eq:lagr-alp} 
 requires the propagation of photons
through external magnetic fields.
In the following we recapitulate the necessary theory for photon-ALP oscillations and summarise the
relevant magnetic fields along the line of sight of an extragalactic $\gamma$-ray source.
A full derivation of the conversion probability which can be found e.g. in refs. \cite{csaki2003,bassan2010,deangelis2011,meyer2013thesis}.

\subsection{Photon-ALP oscillations}
Solving the equation of motion for a mono-chromatic photon-ALP state $\Psi = (A_1,A_2,a)^T$ propagating in a homogeneous $B$ field and plasma,
 where $A_{1,2}$ are the photon polarisation states, leads to a mixing matrix with off-diagonal terms
that induce photon-ALP oscillations \cite{raffelt1988}. 
However, only the magnetic-field component $\mathbf{B}_\perp$ transversal to the propagation direction 
contributes 
to the mixing. Furthermore, only the 
photon polarisation state in the plane spanned by the photon wave vector and the
transversal $B$ field couples to ALPs \cite{raffelt1988,deangelis2011}. 
The off-diagonal term in the mixing matrix is given by  $\Delta_{a\gamma} = g_{a\gamma}B_\perp / 2$
showing the full degeneracy between the magnetic field and the coupling constant (we drop the index of the transversal $B$ field from now on). 
As it turns out, the mixing becomes maximal and independent of energy (the so-called strong mixing regime, SMR)
for $\gamma$-ray energies $E_\mathrm{crit} \lesssim E \lesssim E_\mathrm{max}$,
where
\begin{eqnarray}
E_\mathrm{crit} &=& \frac{|m_a^2 - \omega_\mathrm{pl}^2|}{2g_{a\gamma}B} \sim 2.5\,\mathrm{GeV}\, |m^2_\mathrm{neV} - 1.4\times10^{-3}\, n_{\mathrm{cm}^{-3}}| \,g_{11}^{-1} B_{\mu\mathrm{G}}^{-1},
\label{eq:ecrit}\\
E_\mathrm{max} &=& \frac{90\pi}{7\alpha}\frac{B_\mathrm{cr}^2\,g_{a\gamma}}{B} \sim 2.12\times10^{6}\,\mathrm{GeV}\,g_{11}B_{\mu\mathrm{G}}^{-1},
\label{eq:emax} 
\end{eqnarray}
with $\omega_\mathrm{pl} = 0.037\sqrt{n_{\mathrm{cm}^{-3}}}$\,neV the plasma frequency
which depends on the ambient electron density $n_{\mathrm{cm}^{-3}} = n / \mathrm{cm}^{-3}$,
 and $B_\mathrm{cr}$ the critical magnetic field, $B_\mathrm{cr}\sim4.4\times10^{13}$\,G. 
 For the numerical values we have introduced the notation $B_{\mu\mathrm{G}} = B/(1\mu\mathrm{G})$, $m_\mathrm{neV} = m_a / (1\mathrm{neV}) $, and $g_{11} = g_{a\gamma} \times 10^{11}\,\mathrm{GeV}$.
Above $E_\mathrm{max}$ the oscillations are suppressed due to the QED vacuum polarisation effect.  

The polarisation of $\gamma$-rays cannot be measured with IACTs, so the photon-ALP system is described with the density matrix, $\rho = \Psi \otimes \Psi^\dagger$.
Assuming the propagation direction to lie along the $x_3$ axis and letting $\mathcal{T}$ denote the transfer matrix that solves the equation of motion
for $\Psi$, one finds that the photon survival probability for an initially un-polarised pure photon beam $\rho(x_3 = 0) = \mathrm{diag}(1,1,0) / 2$ is given by
\begin{equation}
P_{\gamma\gamma} = \mathrm{Tr} \left[(\rho_{11} + \rho_{22}) \mathcal T \rho(0) \mathcal T^\dagger \right],
\end{equation}
where $\rho_{ii} = \mathrm{diag}\left(\delta_{i1},\delta_{i2},0\right)$ denotes the polarisation along $x_1$ and $x_2$, respectively. 
In many astrophysical environments the magnetic fields are not homogenous. 
In such cases, the transfer matrix can be split up in $N_d$ domains where in each domain the magnetic field can be treated as constant.
This yields
\begin{equation}
\mathcal{T}(x_{3,N_d},x_{3,1};\psi_{N_d},\ldots,\psi_1;E) = \prod\limits_{i = 1}^{N_d} \mathcal{T}(x_{3,i+1},x_{3,i};\psi_{i};E),
\end{equation}
where $\psi_i$ denotes the angle between the transverse magnetic field and the polarisation state along $x_2$.
By allowing the mixing matrix to include photon absorption, $P_{\gamma\gamma}$ also includes the EBL attenuation 
and reduces to $\exp(-\tau)$ for $g_{11} = 0$. 
Full solutions for $\mathcal T$ are provided, e.g., in refs. \cite{deangelis2011,meyer2013thesis}.

\subsection{Magnetic fields} 
Magnetic fields in numerous environments along the line of sight to a blazar have 
been used to study $\gamma$-ray-ALP oscillations.
The following magnetic fields will be considered here:

\begin{itemize}
\item \textit{Jet magnetic field on pc-scales.}
This field has been studied in terms of photon-ALP conversions in refs. \cite{hochmuth2007,sanchezconde2009,tavecchio2012,mena2013,tavecchio2014,meyer2014}.
The magnetic field at parsec scales close to the VHE emission zone 
is modelled through its toroidal coherent component \cite{begelman1984}, so that $B^\mathrm{jet}(r) = B_0^\mathrm{jet} (r / r_\mathrm{VHE})^{-1}$
and $n^\mathrm{jet}(r) \propto r^{-2}$ under the assumption of equipartition \cite[e.g.][]{osullivan2009}. 
We restrict ourselves to BL-Lac type blazars which are defined through the lack of strong emission or absorption lines in the optical spectrum. 
In contrast to flat spectrum radio quasars (FSRQs), 
these objects do not have a broad line region (BLR) in which high velocity clouds emit broad emission lines \cite[e.g.][]{urry1995}. 
The photon-ALP mixing is most sensitive on $B_0^\mathrm{jet}$ and the distance $r_\mathrm{VHE}$ of the emission site 
to the central black hole \cite{tavecchio2014,meyer2014}.

\item \textit{Jet magnetic field on kpc scales (Lobes).} 
At larger spatial scales,
rotation measures (RM) and synchrotron emission suggest field strengths in the jet of the order of $\mathcal{O}(\mu\mathrm{G})$ 
and up to $100\,\mu$G in hot spots 
in Fanaroff-Riley type II galaxies\footnote{
According to the AGN unification scheme \cite{urry1995}, misaligned flat FSRQs are FR type II galaxies, whereas misaligned BL Lacs are FR type I objects
 \cite[e.g.][]{hardcastle2004,pudritz2012}.
Turbulent magnetic fields have also been detected in the lobes in the misaligned BL Lac Centaurus A \cite{feain2009}.
We neglect any hot spots and 
model the lobe fields with a simple cell-like model in which the magnetic-field strength is constant but the orientation of the field
changes randomly from one cell to the next \cite{tavecchio2014}. 
}.
\item \textit{Intra-cluster magnetic field, ICM.}
Evidence exists that FR type I radio galaxies are often  situated in poor galaxy clusters or groups of galaxies \cite[e.g.][]{longair1979,miller2002}.
Magnetic fields for such environments are deduced from X-ray and radio RM  
\cite[e.g.][]{laing2006,laing2008,guidetti2010,guidetti2011,guidetti2012}. 
The $B$ fields in such environments are commonly modelled with a homogeneous magnetic field with gaussian turbulence
where the energy density follows a power law in wave numbers, $M(k) \propto k^{q}$, between 
the minimum and maximum scales of the turbulence, $k_L = 2\pi / \Lambda_\mathrm{max} \leqslant k \leqslant k_H = 2\pi / \Lambda_\mathrm{min}$. 
The RM of poor environments yield rather flat values of $q > -11 / 3$, where $-11/3$ corresponds to a Kolmogorov turbulence spectrum
\cite{guidetti2011}.
Furthermore, the $B$ field follows the radial dependence of the electron density, $B^\mathrm{ICM}(r) = B^\mathrm{ICM}_0 (n(r) / n_0)^\eta$, where $n(r)$ is 
commonly parametrised with a standard $\beta$ profile\footnote{
The profile is defined through  $n(r) = n_0(1 + r^2/r_c^2)^{-\frac{3}{2}\beta_\mathrm{atm}}$
and usually the central density $n_0$, the core radius $r_c$, and $\beta_\mathrm{atm}$ are determined from X-ray observations \cite[e.g.][]{govoni2004}.
}.
Typical field strengths are of the order of $\mu$G even for almost isolated objects,
where, however, the magnetic field in the radio lobes of the AGN jet could also contribute (see above and \cite{guidetti2012}).
The transversal component for such a field is derived in ref. \cite{meyer2014} which we will also use here.
The random nature of the turbulent fields makes it necessary to simulate a large number of realisations 
and investigate the photon-ALP mixing for each configuration \cite{horns2012icm,hess2013:alps,meyer2014}.

\item \textit{Galactic magnetic field (GMF) of the Milky Way.} 
The importance of the GMF for ALP searches was noted in ref. \cite{simet2008}
and we adopt the coherent component of the model of ref. \cite{jansson2012}, henceforth JF2012,
 which was already used in refs. \cite{horns2012icm,wouters2014,meyer2014}
to study the effect of photon-ALP conversions on VHE $\gamma$-ray spectra.
We discuss the choice of this model in section \ref{sec:disc}.

\end{itemize}
 
 Further magnetic fields exists along the line of sight.
AGN are commonly found in elliptical galaxies \cite{matthews1964} with turbulent magnetic fields 
and coherence lengths of the order of $\mathcal{O}(\mu\mathrm{G})$ and $\mathcal{O}(0.1\,\mathrm{pc})$, respectively \cite{moss1996}.
They are considered as a photon-ALP mixing environment in ref. \cite{tavecchio2012}.
However, only large fields $B \gtrsim 4\,\mu$G would lead to a seizable ALP production and we conservatively neglect their contribution \cite{meyer2014}.
Furthermore, we do not consider mixing in the IGMF, 
which was studied in terms of photon-ALP oscillations in refs.  \cite{mirizzi2007,deangelis2007,mirizzi2009,sanchezconde2009,dominguez2011alps,deangelis2011}. 
Current upper limits for this field are of the order of a few $10^{-9}\,$G 
\cite[e.g.][]{blasi1999} and recently evidence has been reported for RM 
of extragalactic radio sources \cite{neronov2013}, that could be explained with an IGMF field strength of 1\,nG and coherence length of 0.1\,Mpc.
Such rather strong fields could be produced by seed fields from outflows of galactic winds \cite{bertone2006} and 
could cause an imprint of photon-ALP oscillation on $\gamma$-ray spectra \cite[e.g.][]{sanchezconde2009,deangelis2011,dominguez2011alps}.
On the other hand, large scale structure formation simulations suggest smaller values of $10^{-12}\,$G \cite{dolag2005}
and even lower strengths cannot be excluded at present (see, e.g., the review in ref. \cite{durrer2013}).
Due to the large uncertainty in the model parameters for the IGMF, we choose not to consider it here.

Having established the magnetic-field models, we now select promising blazars to investigate the sensitivity 
of CTA to detect a $\gamma$-ray boost induced by photon-ALP oscillations. 

\section{Blazar selection} 
\label{sec:sources}
With the magnetic fields discussed in the previous section we find the following scenario:
 close to the blazar magnetic fields 
 (pc scale jet, lobes, ICM)
  exist in which $\gamma$ rays 
can convert into ALPs. 
The photon-ALP beam propagates towards the observer and photons undergo pair production with EBL photons. 
Once the beam reaches the Milky Way, photons and ALPs mix in the GMF. 
An observer will measure the primary photon flux that survives the EBL attenuation and the ALPs that 
have reconverted into $\gamma$ rays in the GMF.
This latter secondary component will become more pronounced (i) the more ALPs are produced close to the source,
(ii) the stronger the attenuation of the primary photon flux is, and (iii) the more ALPs reconvert into photons in the GMF. 
Point (ii) can be regarded as an effective ``filter" for the primary photons.
The stronger the filter, the easier it is to detect a secondary $\gamma$-ray component. 
Hence, we will consider sources for which observations deep in the optical thick regime are reported in the literature. 
We require that at least one energy bin in the spectrum fulfils $\tau \gtrsim  4$ (with the KD2010 EBL model)
using the central energy of the bin.  
From all extragalactic VHE sources,\footnote{See e.g. \url{http://tevcat.uchicago.edu/} for a catalog.}
 this criterion
 leaves us with the sources 1ES\,0229+200, PKS\,1424+240, and PG\,1553+113 for which we use the lower limit on the redshift $z \geqslant 0.4$ \cite{danforth2010}.
As an additional source, we will consider the distant low frequency synchrotron peaked blazar (LBL) PKS\,0426-380
 which has not been observed with IACTs so far. 
 However, photons at VHE have been detected with the \emph{Fermi}-LAT from this source (see below).
Further source candidates are discussed in section \ref{sec:disc}.
As discussed in the previous section, a number of blazars is associated with poor galaxy clusters, and
we perform a cross correlation between the AGN positions and the GMBCG and WHL galaxy cluster catalogs \cite{hao2010,wen2012},
which cover the redshift ranges $0.1 < z < 0.55$ and $0.05 \leqslant z \leqslant 0.8$, respectively.
In order for the blazar to be associated with one cluster, we demand 
that the angular separation is below the $r_{200}$ radius for the latter catalog\footnote{
The quantity $r_{200}$ defines the radius for which the mean density of the cluster is 200 times the critical density of the universe \cite{wen2012}. 
It is not provided in the GMBCG catalog.}
and 2\,Mpc for the former (constituting a conservative, i.e. large estimate for $r_{200}$)
 and that the redshift distance obeys the inequality $\Delta z = |z - z_\mathrm{cl}| \leqslant 0.05$, where $z_\mathrm{cl}$ is the photometric redshift of the cluster.
These values are used in the above catalogs to cross correlate their identified clusters with other catalogs.

\paragraph{1ES\,0229+200} This ultra high synchrotron peaked blazar at a redshift $z = 0.1396$ \cite{woo2005}
has been observed with H.E.S.S. for 41.8 hours \cite{1es0229hess2007} and with VERITAS for 54.3 hours \cite{1es0229veritas2013}.
The observations revealed a hard observed $\gamma$-ray spectrum with $\Gamma_\mathrm{obs} = 2.50\pm0.19_\mathrm{stat}\pm0.10_\mathrm{sys} $ \cite{1es0229hess2007}. No sign for variability on any time scale is present in H.E.S.S. data whereas  
evidence is found for flux variations on a yearly time scale in the VERITAS observations. 
The H.E.S.S. spectrum extends beyond 11\,TeV or an optical depth of $\tau \sim 4.4$. 
For the simulations conducted here (see section \ref{sec:method} and \ref{sec:results}), we will use the H.E.S.S. spectrum and assume the same observation time.

A multi-wavelength study including the VERITAS data and a one-zone synchrotron-self-Compton (SSC) model 
results in a best fit of the magnetic field in the emission zone between $7.5\times10^{-4}$\,G and $2.6\times10^{-3}$\,G, 
a Doppler factor\footnote{
The Doppler factor is defined through $\delta = \left[\Gamma_\mathrm{L}(1 - \beta_\mathrm{j} \cos\theta)\right]^{-1}$,
where $\Gamma_\mathrm{L}$ is the bulk Lorentz factor of the jet, $\beta_\mathrm{j}$ the corresponding velocity (in units of $c$)
and $\theta$ is the angle between the jet axis and the line of sight. 
} $56.4 \leqslant \delta \leqslant 100$ and a radius of $5.8\times10^{15}$\,cm of the VHE emitting plasma blob \cite{1es0229veritas2013}.
The comparatively small field strength will unlikely lead to a seizable ALP production \cite{tavecchio2014,meyer2014}
 and we neglect its contribution here.
We find that the blazar is within a distance of  $0.72\,\mathrm{Mpc} < r_{200}$ from the cluster WHL\,22793 at a redshift difference 
of $\Delta z = 8\times10^{-4}$. The cluster has 9 galaxies associated with it.
The $B$ field in this environment is unknown and one has to rely on observations of close-by radio galaxies 
found in similar surroundings for which the magnetic field is known.  
For definiteness, we adapt the magnetic field as found around the FR I radio galaxy 3C\,449 that is also located in a group of galaxies \cite{guidetti2010}. 
With the electron density 
adopted from ref. \cite{croston2008}, $n_0 = 3.7\times10^{-3}\,\mathrm{cm}^{-3}$, $r_c \sim 20\,$kpc, and $\beta_\mathrm{atm} = 0.42 \pm 0.05$,  the authors of \cite{guidetti2010} find a central magnetic field of $B_0 = (3.5\pm1.2)\,\mu\mathrm{G}$ 
that follows the electron density with a broken power law. Also the turbulence spectrum $M(k)$ is best fitted with a broken power law. 
For simplicity, will assume single power laws, i.e. $\langle B(r) \rangle  = B_0(n_e(r) / n_0)^\eta$ with $\eta = 1$ 
and the turbulence spectrum with index  $q = 2.57$, the average of the two indices found in ref. \cite{guidetti2010},
with minimum and maximum wave numbers $k_L =  0.015\,\mathrm{kpc}^{-1}$ and $k_H = 5\,\mathrm{kpc}^{-1}$, respectively. Beyond a distance of 100\,kpc we set the magnetic field to zero.  

\paragraph{PG\,1553+113} The variable high synchrotron peaked BL Lac (HBL) has been observed with
 H.E.S.S. for 7.6 hours in 2005 and 17.2 hours in 2006 \cite{pg1553hess2005,pg1553hess2008}, with MAGIC for 9.5 hours in 2006 and 7.2 hours in 2008 \cite{pg1553magic2007,pg1553magic2010}, and with VERITAS for 50 hours between 2010 to 2011 \cite{pg1553veritas2010}. 
Using the lower limit $z \geqslant 0.4$ \cite{danforth2010},
 the H.E.S.S. observations extend to an optical depth $\geqslant 3.95$ for the highest energy data point at $0.949$\,TeV. 
In 2012, MAGIC observed the source in a flaring state lasting for several days reaching up to 100\,\% of the Crab nebula flux above 100 GeV 
in an observation campaign spanning a total of 17.4 hours \cite{pg1553magic2013}. 
We will use the H.E.S.S. observation of the source since the MAGIC observations do not extend to $\tau \sim 4$.
To emulate the flare spectrum the flux is upscaled by a factor of 3.58 \cite{meyer2014}.
 A 20 hour observation will be simulated. 

No  broadband modelling of the spectral energy distribution (SED) is available for the flaring state of this source and hence no information about the
magnetic field on pc scales and on the position of the VHE emitting zone is available. 
Thus, we do not include any contribution of this magnetic field to the photon-ALP mixing.
A cross correlation with the galaxy cluster catalogs reveals that the source is within $1.39\,$Mpc of a poor cluster with 
8 member galaxies and ID GMBCG 587742629068931733. The redshift difference is $\Delta z = 0.035$, assuming $z = 0.4$. 
We adopt the scenario of photon-ALP
mixing in the cluster environment, taking the same magnetic field and electron density as for 1ES\,0229+200 with 
the exception of the $B$-field strength which we conservatively fix to $1\,\mu$G. 

\paragraph{PKS\,1424+240} A lower limit on the redshift of $z \geqslant 0.6035$ of this intermediate peaked synchrotron blazar (IBL) was recently determined \cite{furniss2013} making it the most distant VHE source today. 
It has been observed by MAGIC for 33.6 hours between 2009 and 2011 \cite{pks1424magic2014} and VERITAS in two campaigns \cite{pks1424veritas2010,pks1424veritas2014} for 28.5 and 67 hours. 
The second deep VERITAS observation campaign results in a significant detection beyond 500\,GeV or $\tau = 4.1$ using the lower limit on $z$.
This spectrum will be used here for the simulations together with the same observation time. 

An SSC fit (for $z = 0.7$) to multi-wavelength data including the 2009 VERITAS observations gives the best-fit values of
$B = 0.14\,$G, $R = 5\times10^{16}$\,cm, and $\delta = 60$ \cite{pks1424veritas2010}. 
Including radio data and the MAGIC observations, the authors of ref. \cite{pks1424magic2014} conclude that 
a one-zone SSC model does not suffice to satisfactorily describe the broadband SED and the radio morphology. 
Instead, they apply a two-zone SSC model which gives $B = 0.033\,$G, $R = 4.8\times10^{16}$\,cm, and $\delta = 30$
for the zone responsible for the VHE emission.
Assuming an angle between the jet axis and the line of sight of $1^\circ$ and a simple conical jet geometry yields a 
distance of $r_\mathrm{VHE} \sim 0.06$\,pc of the emission zone to the central black hole. 
The IBL cannot be associated with any cluster of the WHL catalog (it is outside the redshift range of the GMBC catalog)
 within the defined search criteria, even though its position is covered in the SDSS
survey. The WHL catalog is not complete for $z > 0.42$ \cite{wen2012} and it is therefore possible that the cluster is not detected 
or does not fulfil the minimum requirement of at least 8 galaxies within $r_\mathrm{200}$. 
Thus, we will only assume a photon-ALP mixing within the BL Lac jet with the values of the two-zone SSC model fit.
We set the maximum scale of the coherent magnetic field in the jet to 1\,kpc (and to zero above) and the angle between the transversal magnetic field 
and the propagation direction to $45^\circ$ (as done in \cite{tavecchio2012}).

\paragraph{PKS\,0426-380} This distant LBL object (or possibly FSRQ, \cite{tanaka2013}) is located at a redshift of $z = 1.111$ \cite{heidt2004}. 
No measurement of the source with any IACT is reported in the literature so far. Nevertheless, it has been observed with the \emph{Fermi}-LAT
and a dedicated analysis has revealed two photons with energies above 100\,GeV ($\tau > 1$) that can be associated with the source with high significance \cite{tanaka2013}. 
The blazar is variable and the averaged observed spectrum during the flaring episodes 
can be described with a power law with index $\Gamma_\mathrm{obs} = 2.72\pm0.17$ above $8$\,GeV with mild evidence for a spectral flattening 
above 10\,GeV \cite{tanaka2013}. 
This source is included in order to test the sensitivity to a $\gamma$-ray boost at low energies. 
The large redshift leads to an energy for which $\tau = 4$ of $E_{\tau = 4} \sim 235$\,GeV.
Thus, the attenuation is sensitive to the UV and optical part of the EBL spectrum.

No information on the magnetic field in the jet from, e.g., SSC modelling is available at this point. 
Furthermore, the cluster catalogs do not cover the required redshift range (its position is anyway not included in the SDSS).
Therefore, we will tentatively assume a photon-ALP mixing within the lobes of the AGN jet as suggested in ref. \cite{tavecchio2014}.
As discussed in section \ref{sec:alp}, evidence for magnetic fields in lobes is deduced from RM and the observation of synchrotron emission from the lobes.
Here, we model the lobe magnetic field with a simple cell-like structure, where each cell has a length of $L_\mathrm{coh} = 10$\,kpc. 
Motivated by the values deduced from the lobes of Centaurus A \cite{feain2009}, the field strength is assumed to be constant with $B = 1\,\mu$G and to extend over 100\,kpc.

\begin{figure}[tb]
\centering
\includegraphics[width = .7 \linewidth]{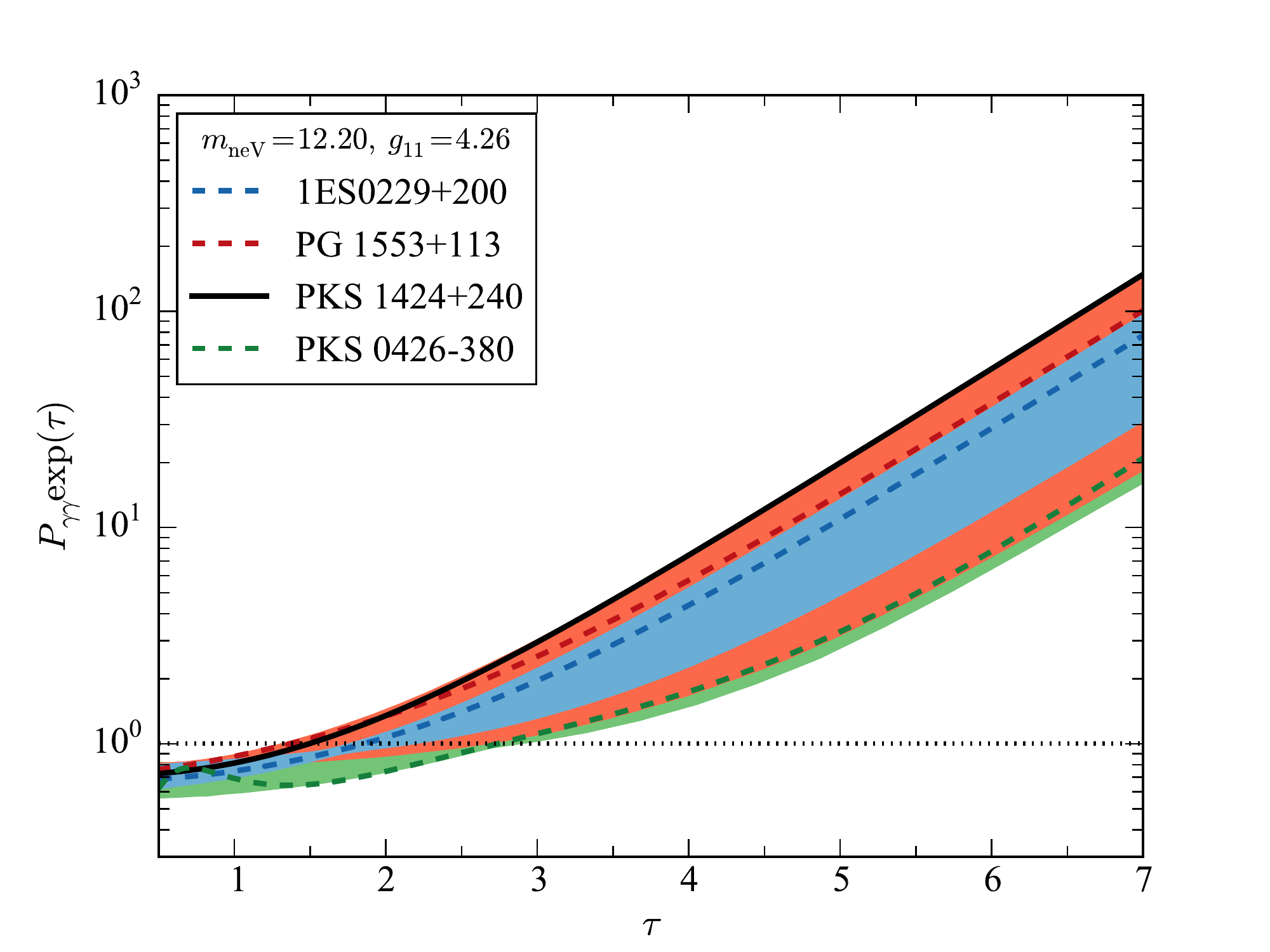}
\caption{Boost of the photon flux due to photon-ALP mixing versus the optical depth for one particular choice of $m_a$ and $g_{a\gamma}$. 
For all blazars except PKS\,1424+240, the magnetic fields are random and the coloured bands show
the 68\,\% envelope around the median for 1000 realisations of the $B$ fields. The dashed lines correspond to 
one random realisation. For PKS\,1424+240, the mixing occurs in the coherent magnetic field of the BL Lac jet.}
\label{fig:boost}
\end{figure}

The above blazars used for this study are summarised with their sky coordinates, magnetic-field scenarios, and assumed observation times in table \ref{tab:srcs}. 
For the proposed magnetic fields, we show the boost, i.e., the ratio between the photon survival probability with ALPs, $P_{\gamma\gamma}$, and without ALPs, $\exp(-\tau)$, for each of the sources in figure \ref{fig:boost}.
In the figure, the photon-ALP coupling is set to $g_{11} = 4.26$, 
compatible with the upper limit of $g_{11} = 6.6$ derived from the observations of globular clusters \cite{ayala2014}. 
At low values of $\tau$, the photon-ALP mixing leads to a drop in the $\gamma$-ray flux. 
Above $\tau \sim 4$, however, the boost exceeds $\sim 2$ and reaches factors up to $\sim 10$ as the primary component becomes stronger attenuated while
the reconverted ALPs give a constant contribution to the total flux.
The figure underlines that observations at large value of $\tau$ are best suited 
to detect a secondary $\gamma$-ray component. 
In the following, CTA observations of these sources will be simulated in order to determine the sensitivity to 
detect ALP induced boosts.

\begin{table}[tb]
\centering
\begin{footnotesize}
\begin{tabular}{|l|cccc|}
\hline
\multirow{2}{*}{Parameter} & \multicolumn{4}{|c|}{Sources} \\
{} &1ES\,0229+200 & PG\,1553+113 & PKS\,1424+240 & PKS\,0426-380\\
\hline
\hline
Redshift $z$ & 0.139 & $\geqslant 0.4$ & $\geqslant 0.6035$ & 1.11 \\
R.A (deg)& 38.20192 & 238.92933 & 216.75162 & 67.16842 \\
Dec. (deg) & 20.28783 &  11.19011 & 23.8 & $-37.93878$ \\
\hline
$B$-field scenario & ICM + GMF & ICM + GMF & Jet + GMF & Lobes + GMF \\
$B_0\,(\mu\mathrm{G})$ & 3.5& 1 & $3.3\times10^4$ & 1 \\
$r_\mathrm{VHE}\, (\mathrm{pc})$ & -- & -- & 0.057 & -- \\
$r_\mathrm{max}\, (\mathrm{kpc})$ & 100 & 100 & 1 & 100 \\
$\delta_\mathrm{D}$ & -- & -- & 30 & -- \\
$\eta$ & 1 & 1 & -- & -- \\
$\beta_\mathrm{atm}$ & 0.42 & 0.42 & -- & -- \\
$r_c\, (\mathrm{kpc})$ & 19.33 & 19.33 & -- & -- \\
$n_0\, (\mathrm{cm}^{-3})$ & $3.7 \times 10^{-3}$ & $3.7 \times 10^{-3}$ & $10^4$ & $10^{-3}$ \\ 
$L_\mathrm{coh}\, (\mathrm{kpc})$ & -- & -- & -- & 10 \\
$q$ & -2.53 & -2.53 & -- & -- \\
$k_L\,(\mathrm{kpc}^{-1})$ & 0.015 & 0.015 & -- & -- \\
$k_H\,(\mathrm{kpc}^{-1})$ & 5 & 5 & -- & -- \\
\hline
$T_\mathrm{obs}$ (hours) & 41 & 20 & 67 & 70 \\ 
ref. for spectrum & \cite{1es0229hess2007} & \cite{pg1553hess2005} & \cite{pks1424veritas2014} & \cite{tanaka2013} \\
\hline
\end{tabular}
\end{footnotesize}
\caption{Blazars used for this study together with the assumed magnetic fields close to the source.
 The sky coordinates are taken from the Roma BZCAT catalog \cite{bzcat}. The last row gives
the reference from which the observed spectrum is taken that is used as an input for the simulations.
 See text for further details.}
\label{tab:srcs}
\end{table}

\section{Method}
\label{sec:method} 
We proceed by introducing the CTA simulations and the statistical procedure to assess the sensitivity to detect a 
$\gamma$-ray boost. 
We closely follow the method outlined in ref. \cite{meyer2014} to which we refer the reader for further details. 

\begin{figure}
\centering
\includegraphics[width = .49 \linewidth]{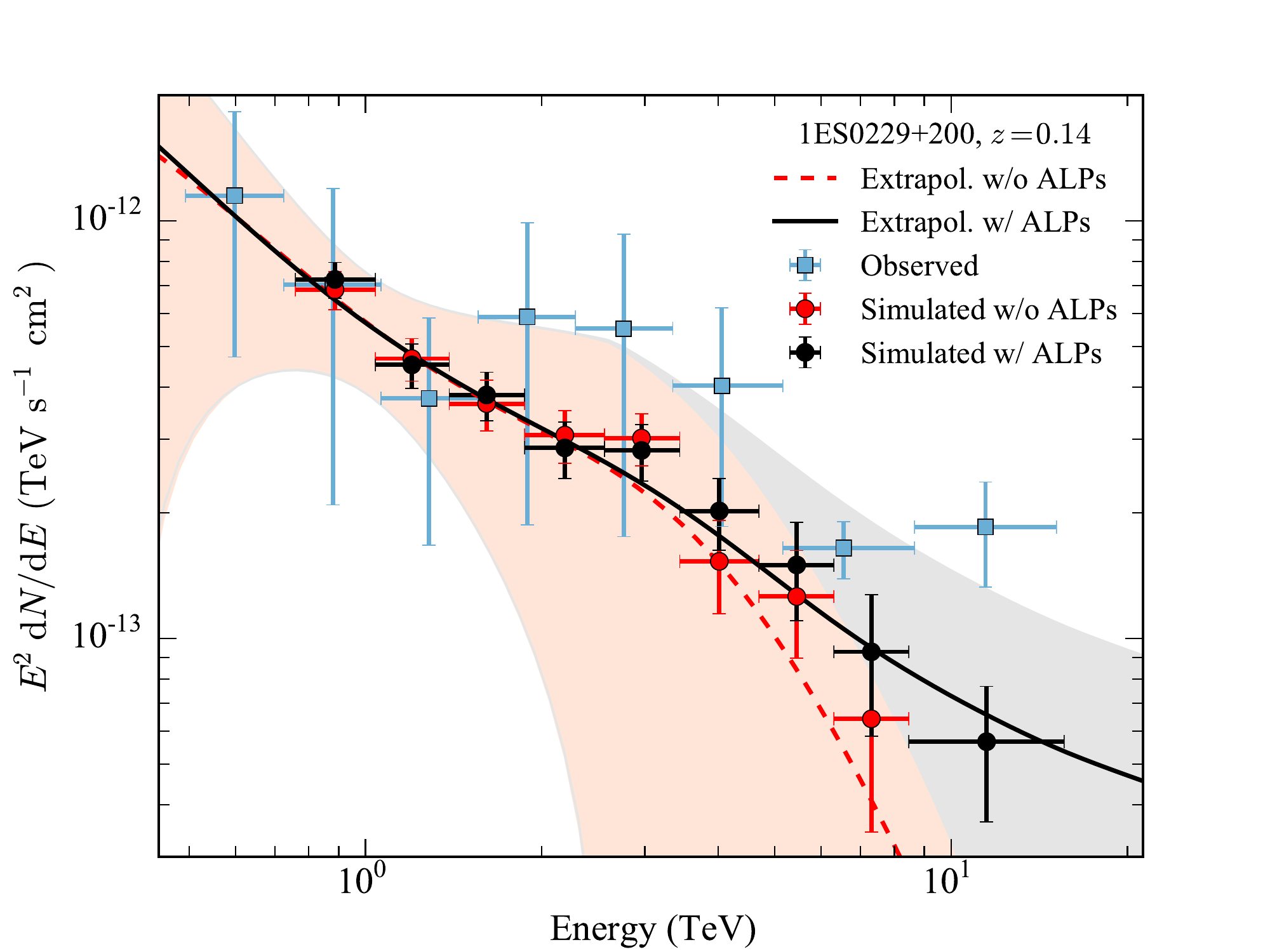}
\includegraphics[width = .49 \linewidth]{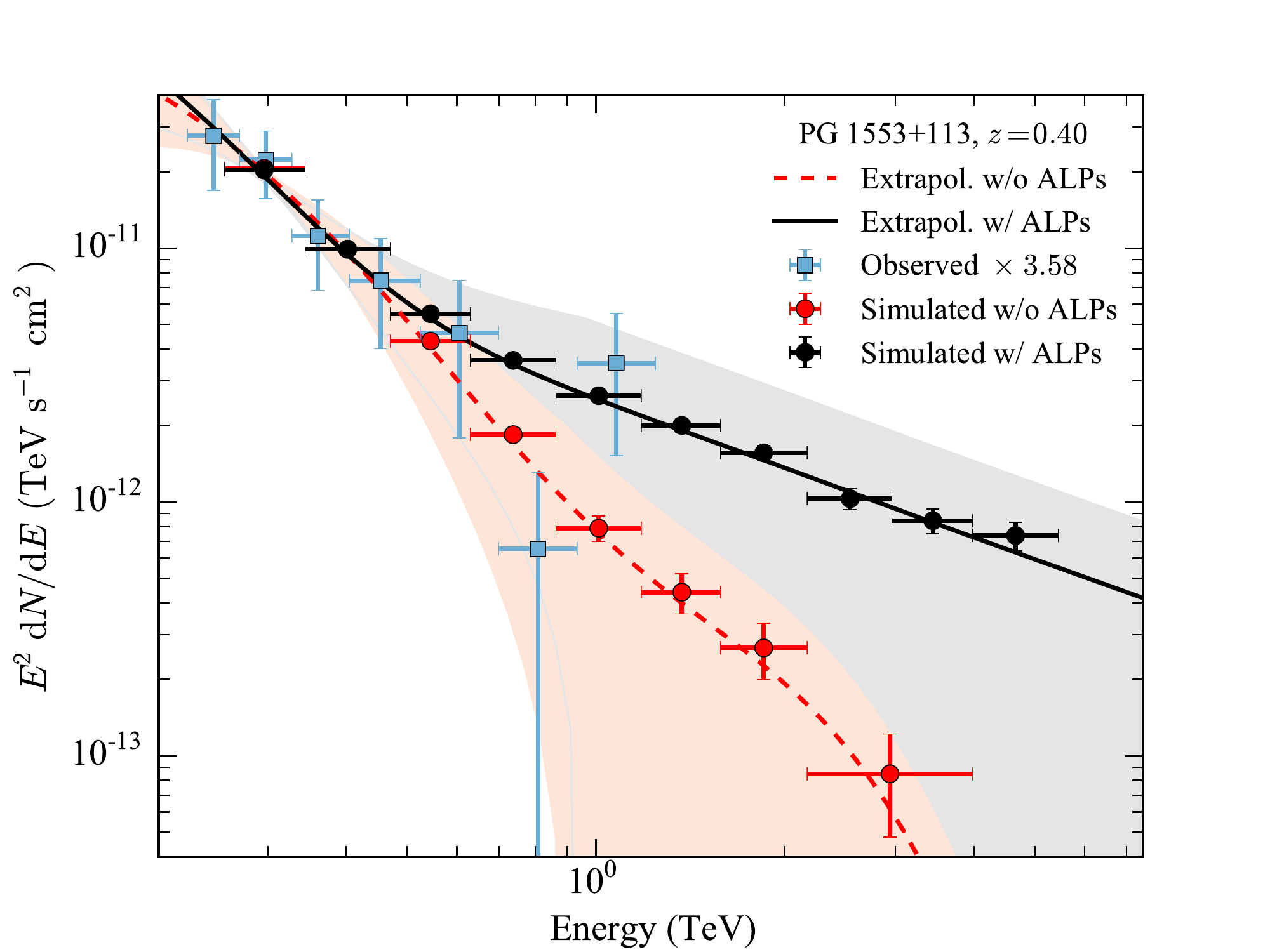}
\includegraphics[width = .49 \linewidth]{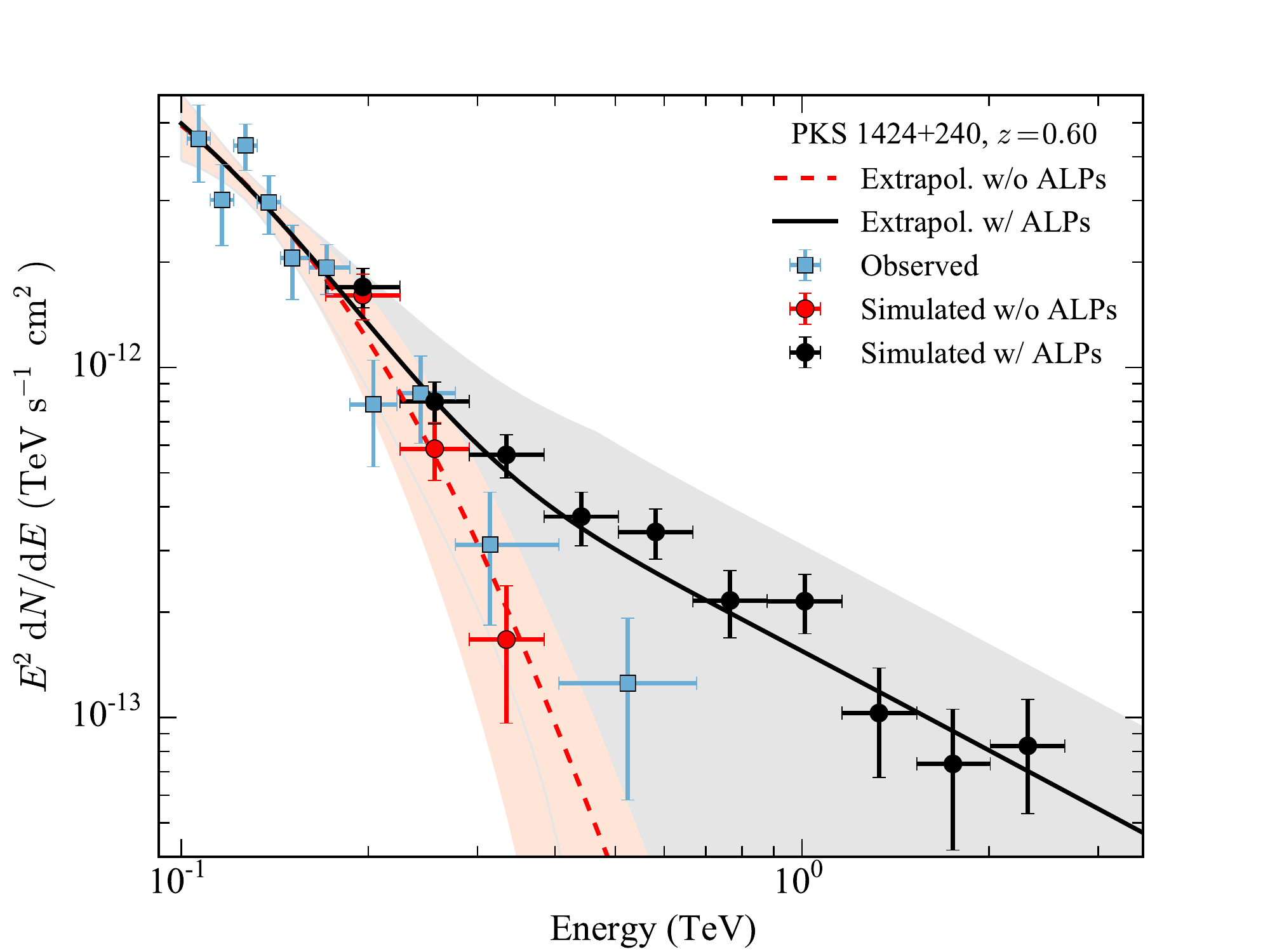}
\includegraphics[width = .49 \linewidth]{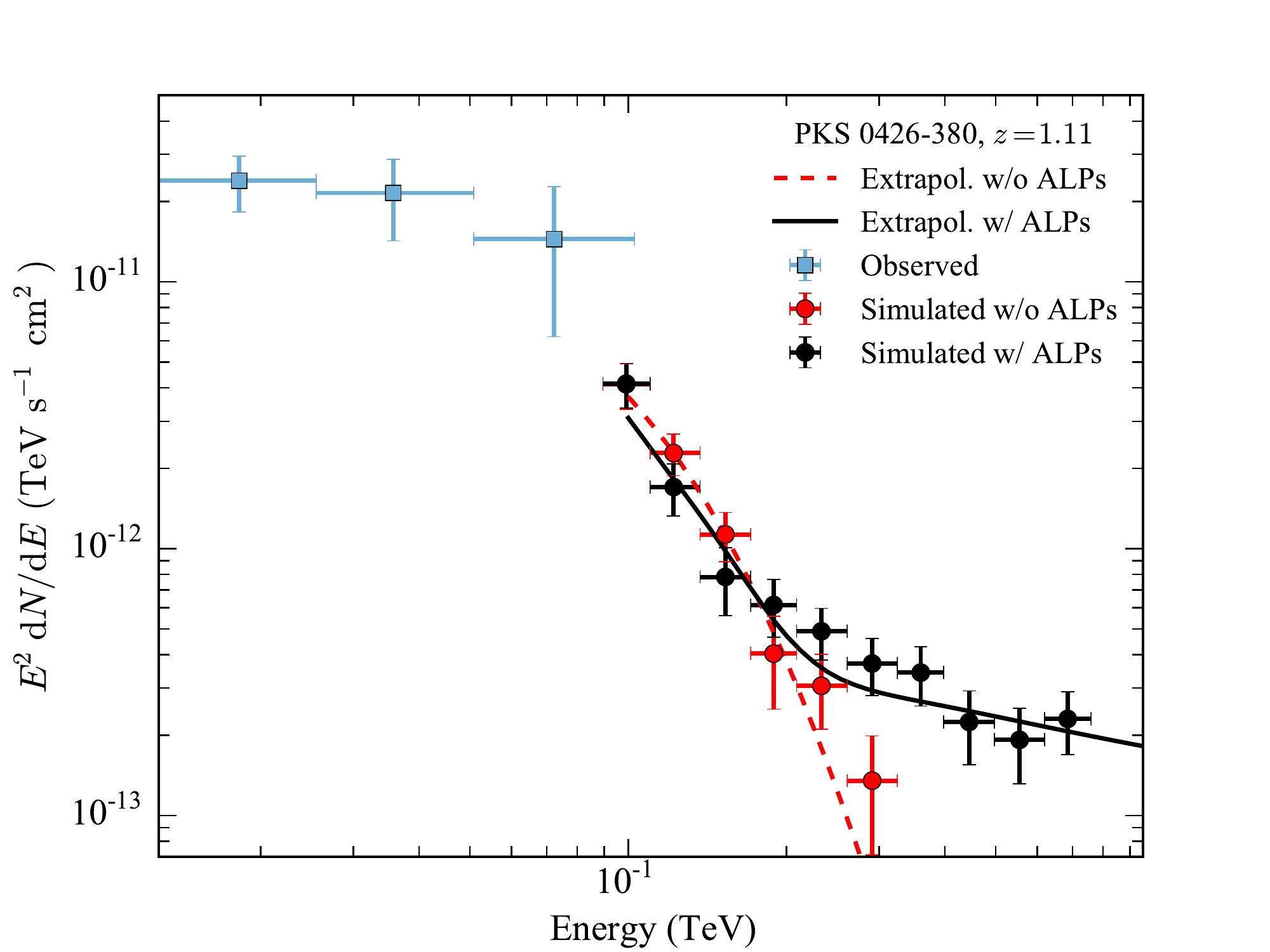}
\caption{Simulated spectra for the blazars selected in section \ref{sec:sources}
for $\tau \geqslant 1$ with ALPs (black bullets and solid lines) and without (red bullets and dashed lines). 
The ALP parameters are set to $m_\mathrm{neV} = 12.2$, $g_{11} = 4.3$.
The envelope shapes show the uncertainty of the determined intrinsic spectrum ($1\,\sigma$ fit uncertainty). 
Observed spectra are shown as blue squares. }
\label{fig:specs}
\end{figure}

The $\gamma$-ray spectrum observed at Earth, $\phi_0(E)$, serves as an input for the simulation of the observation
and is given by the product of the intrinsic source spectrum $\phi(E)$ and the photon survival probability $P_{\gamma\gamma}$. 
The latter quantity is determined by the ALP and magnetic-field parameters and reduces to 
the standard EBL absorption $\exp(-\tau)$ in the case of $g_{11} = 0$.
The intrinsic source spectrum is unknown and we estimate $\phi(E)$ 
for the three sources that are already observed with IACTs  in the following way. 
We fit a power law $\phi_\mathrm{obs}(E) = N_\mathrm{obs} (E / E_0)^{-\Gamma_\mathrm{obs}}$ to the observed spectral points
 from the references listed in table \ref{tab:srcs}\footnote{All observed spectra considered here are satisfactorily described with a power law.}.
The observed spectrum is corrected for absorption and possible ALP effects in each energy bin $\Delta E$
 with $1/\langle P_{\gamma\gamma}\rangle$, where
\begin{equation}
\langle P_{\gamma\gamma}\rangle = \frac{\int\limits_{\Delta E}\mathrm d E\, P_{\gamma\gamma}\, \phi_\mathrm{obs}(E)}{\int\limits_{\Delta E}\mathrm d E \,
 \phi_\mathrm{obs}(E)}.
\end{equation}
We determine the intrinsic spectrum, $\phi(E) = N(E/E_0)^{-\Gamma}$, from yet another power-law fit to the absorption corrected data points.
In order to be as independent as possible from the exact shape of $P_{\gamma\gamma}$ only data points in the optical
thin regime, i.e., $\tau < 1$ should be included. However, for a reasonable determination of $\phi(E)$ we require at least 4 data points in the fit. 
Thus, for 1ES\,0229+200 (PG\,1553+113) data points up to $\tau \sim 1.57$ (2.17) are taken into account. 
For PKS\,1424+240, there are 5 data points with $\tau < 1$ which are included in the fit.
Thanks to the envisaged energy threshold of CTA around $\sim30\,$GeV it will be possible to determine the intrinsic spectrum at low energies
with much higher confidence in future observations. 
In the case of PKS\,0426-380, where no IACT measurement is available, we simply set $\phi(E)$ to the \emph{Fermi}-LAT spectrum
above 8\,GeV but assume a slightly harder index, $\Gamma = 2.3$, since the observed spectrum already suffers from attenuation.
The $\gamma$-ray spectrum at Earth is then  $\phi_0(E) =P_{\gamma\gamma} \phi(E) $. 
We assume that $\phi(E)$ extends to $\tau = 11$ (which corresponds to an extrapolation about an order of magnitude in energy) 
and set it to zero for higher energies\footnote{
We assume that the intrinsic spectra do not harden with energy.
}. In section \ref{sec:disc} we will examine how the results change if a logarithmic parabola\footnote{
The logarithmic parabola is defined through $\phi(E) = N(E/ E_0)^{-(\alpha + \beta\ln(E / E_0))}$.
}
with negative curvature, i.e. $\beta > 0$, is used instead of a power law.
The spectrum is folded with the IRF of CTA which depends on the true ($E$) and reconstructed energy ($E'$) 
and consists of the point spread function (which we neglect in the following), 
the effective area, $A_\mathrm{eff}(E)$, and the energy dispersion, $D_E(E',E)$. 
It is subsequently multiplied with the observation time (cf. table \ref{tab:srcs}). 
This gives the number of expected counts in each energy bin $\Delta E'_i$,
$\mu_i$, $i = 1,\ldots,n$, 
\begin{equation}
\mu_i = T_\mathrm{obs} \int\limits_{\Delta E'_i}  \mathrm{d}E' \int\mathrm{d} E~ D_E(E,E') ~ A_\mathrm{eff}(E) \phi_0(E),
\end{equation}
which depend on the ALP parameters $(g_{a\gamma},m_a)$ as well as on a number of nuisance parameters: 
The parameters of the intrinsic spectrum, $N,\Gamma$, the magnetic-field model (and particular realisation in case of random fields), 
and the EBL. 
We use the IRF determined in dedicated simulations for  the preliminary array E configuration of CTA, which constitutes a compromise
in source sensitivity at low and high energies \cite{bernloehr2013}. With this set up, it will be possible to  determine the intrinsic spectrum and 
simultaneously measure the spectrum in the optical thick regime.  
A constant zenith angle of $20^\circ$ and a ratio between source and off-source exposure of $\alpha = 0.2$ are assumed
and the number of expected background events in each energy bin, $b_i$, is also obtained from simulations \cite{bernloehr2013}.
Examples for the simulated spectra with and without an ALP contribution are shown in figure \ref{fig:specs}. 
The individual flux points are derived by integrating $\phi_0$ over each energy bin and weighting the bin with the ratio 
of observed and ideally expected excess counts,
\begin{equation}
\left(\frac{\mathrm{d}N}{\mathrm{d}E}\right)_i = \frac{1}{\Delta E'_i}\frac{N_{\mathrm{excess},i}}{\mu_i} \int\limits_{\Delta E'_i} \mathrm d E~ \phi_0(E),
\end{equation}
where $N_{\mathrm{excess},i} = N_{\mathrm{ON,i}} - \alpha N_{\mathrm{OFF},i}$ with the random numbers $N_{\mathrm{ON,i}}$ and $N_{\mathrm{OFF},i}$
 for the source (ON) and background region (OFF) drawn from Poisson distributions with means $\mu_i + b_i$ and $b_i / \alpha$,
respectively. 
 A spectral flattening in the ALP case is observed for all blazars if ALPs are included. 

Following ref. \cite{meyer2014}, we calculate the sensitivity of CTA to detect an ALP induced spectral hardening by means of the likelihood ratio test and make use of the Asimov data set \cite{cowan2011}.
The likelihood $\mathcal L$ in the $i$-th energy bin is 
given as the product of the Poissonian probability mass functions of the events from the ON and OFF region, 
\begin{equation}
\mathcal L(\mu_i,b_i;\alpha | N_{\mathrm{ON},i}, N_{\mathrm{OFF},i}) = \mathrm{Pois}(N_{\mathrm{ON},i} | \mu_i + b_i)\,\mathrm{Pois}(N_{\mathrm{OFF},i} | b_i / \alpha).
\end{equation} 
For an  Asimov data set the number of counts is equal to the expected value, $N_\mathrm{ON} = \mu + b$ and $N_\mathrm{OFF} = b / \alpha$ in each energy bin, making the expected values $\mu$ and $b$ the maximum likelihood estimators. 
The sensitivity to exclude the hypothesis of having no photon-ALP mixing ($\tilde\mu = \mu (g_{a\gamma} = 0)$) 
given an observation including an ALP contribution is then evaluated with the test statistic
\begin{equation}
\mathit{TS}_A = -2\sum\limits_{\substack{i,j\\\tau(E_{ij},z_j)\, > \,2 \\ S_{ij}\, > \,2\,\sigma}}
\ln\left(\frac{
\mathcal L (\tilde\mu_{ij},\widehat{\widehat{b}}_{ij}(\tilde{\mu}_{ij}) ;\alpha | \mu_{ij} + b_{ij}, b_{ij} / \alpha)
}
{
\mathcal L (\mu_{ij},b_{ij} ;\alpha | \mu_{ij} + b_{ij}, b_{ij} / \alpha)
}\right).\label{eqn:loglRatio}
\end{equation}
The sum runs over all sources $j = 1,\ldots,4$ and energy bins $i$ that are detected with a significance above $2\,\sigma$ (using eq. (17) of ref. \cite{lima1983}) and for which the central energy $E_{ij}$ fulfils $\tau(E_{ij},z_j) > 2$. 
Bins with a lower significance will be joined and included if their combined sensitivity is above $2\,\sigma$.
In the numerator, $\widehat{\widehat{b}}_{ij}$ indicates 
the maximisation of the likelihood for fixed $\tilde\mu_{ij}$. 
The sensitivity for a single source is evaluated by omitting the sum over $j$.
In order to convert a $\mathit{TS}_A$ (the subscript $A$ denotes the use of the Asimov data)
into a significance with which 
one can exclude the no-ALP hypothesis, 
the underlying distribution of the test statistic (the null distribution) has to be known. 
For the same reasons outlined in ref. \cite{meyer2014},
we determine the null distribution of the $\mathit{TS}_A$ values with Monte-Carlo simulations and find that it can roughly be described
with a $\chi^2$ distribution with 7 degrees of freedom (d.o.f.)\footnote{
In ref. \cite{meyer2014} the null distribution was best described with 6 d.o.f. The difference is due to the different IRF applied here.
}, $f(\chi^2,7)$. 
We define the sensitivity by the significance $\alpha_0$ to exclude the $g_{11} = 0$ hypothesis by
\begin{equation}
1 - \alpha_0 = \int\limits_{\mathit{TS}_A}^\infty f(\chi^2,7) \mathrm{d}\chi^2.
\end{equation}
The $\mathit{TS}_A$ values are calculated over a grid in the ALP parameter space $(m_a,g_{a\gamma})$. 
We test ALP masses in the range $m_\mathrm{neV} \in [0.1,300]$ and couplings between $g_{11} \in [0.3,7]$. 
For the maximum ALP mass tested, the energies of all bins of the simulated spectra should fall below the 
critical energy defined in eq. \eqref{eq:ecrit} and no photon-ALP mixing is expected. 
For the minimal mass value, all data points lie within the strong mixing regime 
and lower masses will not change the results.
The values for the coupling are motivated, on the one hand, by the upper bound derived from the observations of globular clusters limiting 
$g_{11} < 6.6$ \cite{ayala2014} and, on the other hand, by a negligible ALP production at $g_{11} = 0.3$ \cite{meyer2014}. 
We choose a logarithmic spacing of the grid and $21\times20$ grid points.
In the cases where the magnetic field is turbulent, 1000 random realisations of the $B$ fields are simulated giving the 
same number of $\mathit{TS}_A$ values for each grid point. 
We will present the results in the next section for 
$B$-field configurations which result in a  $\mathit{TS}_A$ value 
 that corresponds to a certain quantile $Q$ of the cumulative distribution function (CDF) of all $\mathit{TS}_A$ values, i.e. $\mathrm{CDF}(\mathit{TS}_A)  = Q$.
For example, for the $TS_A$ value for which $Q = 0.95$,
 95\,\% of the $B$-field realisations result in a weaker rejection of the no-ALP hypothesis, 
and hence this particular $B$ field can be considered as an optimistic configuration in terms of photon-ALP mixing. 
We will consider the quantiles $Q = 0.05,0.5$, and $Q = 0.95$. 

\section{Results} 
\label{sec:results}
\begin{figure}
\centering
\includegraphics[width = .9 \linewidth]{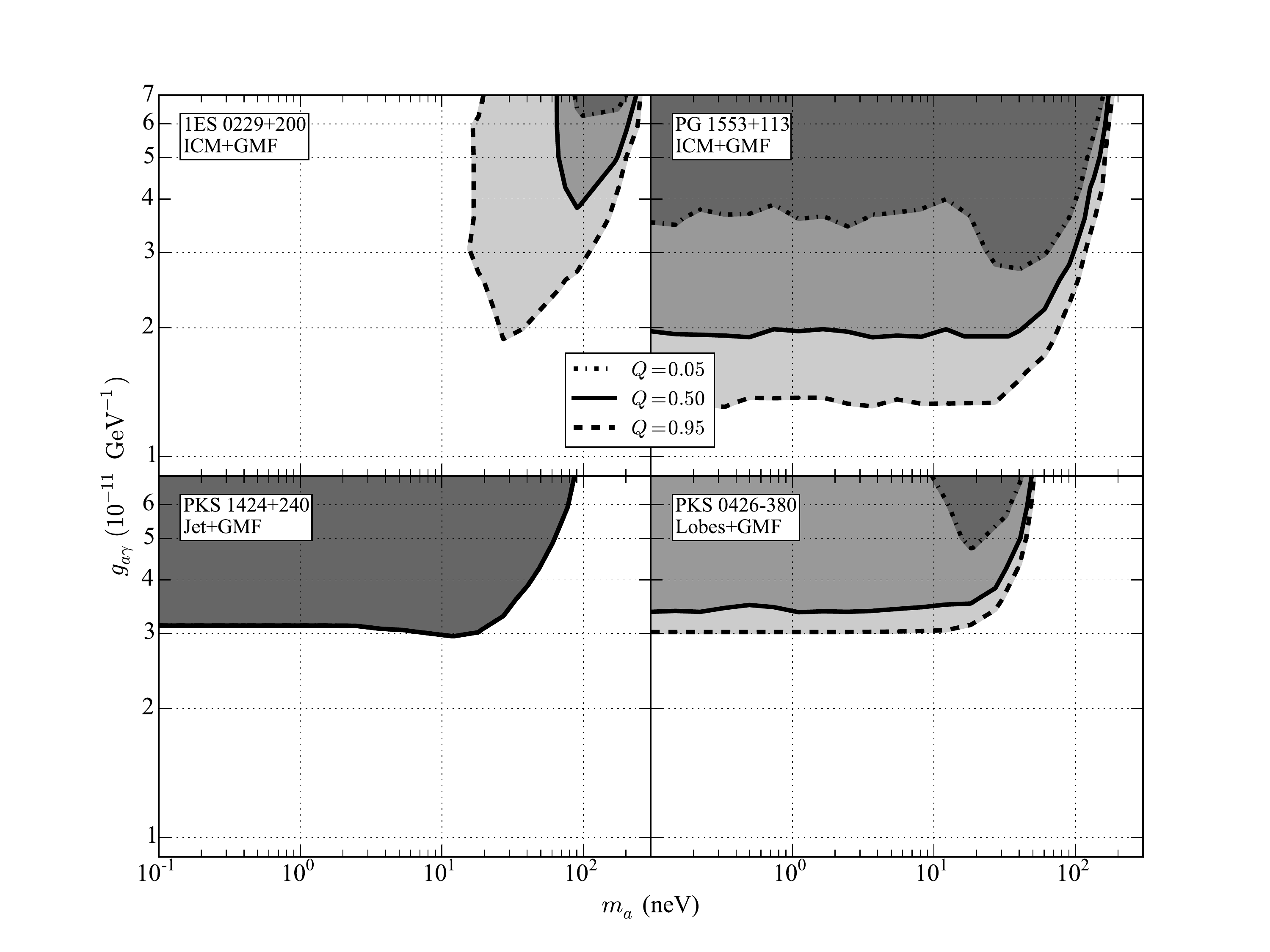}
\caption{Contour lines in the ALP parameter plane above which the no-ALP hypothesis is in tension 
with the simulated observations at the $3\,\sigma$ level. The different panels
show the results for each considered source. In case the source magnetic field is 
modelled with a random field, the contour lines for three different values of the quantile $Q$ are shown.}
\label{fig:single_src}
\end{figure}

The 3\,$\sigma$ sensitivity for an ALP detection for each considered source is shown in figure \ref{fig:single_src} in the $(m_a, g_{a\gamma})$ plane.
We assume the magnetic field models as described in section \ref{sec:sources} together with the JF2012 GMF model and 
the KD2010 EBL model. 
For turbulent magnetic fields, the contour lines for different quantiles $Q$ as defined in the previous section are displayed. 
Above the contours lines, the no-ALP hypothesis is in tension with the observations above the  $3\,\sigma$ level.
With the exception of 1ES\,0229+200, the general trend is as expected. 
The detection sensitivity is independent of the ALP mass until the first energy bins with $\tau > 2$ fall outside the SMR. 
Outside the SMR, the ALP effect is reduced what can be compensated by higher photon-ALP couplings. 
This results in an upturn of the sensitivity contours. 
The contour lines for pessimistic random magnetic field realisations ($Q = 0.05$, dash-dotted lines) for 
PG\,1553+113 (top-right panel of figure \ref{fig:single_src}) show a slightly different behaviour. 
Towards higher masses, the sensitivity actually increases. 
This is due to the spectral irregularities that occur around $E_\mathrm{crit}$
 which have been used to place limits on the photon-ALP coupling with H.E.S.S. observations of PKS\,2155-304 \cite{hess2013:alps}.
It is also the regime where a detection of ALPs from 1ES\,0229+200 (top-left panel of figure \ref{fig:single_src}) would be possible. 
ALPs would not be detected from a boost in the $\gamma$-ray flux alone from this HBL. 
The reason for this lies in the derivation of the intrinsic spectrum
(cf. section \ref{sec:method}). 
The power-law fit to the first four data points of the absorption corrected observed spectrum
results in a soft power-law index and consequently a low flux of the source at high optical depths (see the top-left panel of figure \ref{fig:specs}).
Without ALPs, the intrinsic spectrum is found to be $\Gamma = 2.31 \pm 0.99$ with the procedure described in 
section \ref{sec:method}.
For comparison, the spectrum measured with the \emph{Fermi}-LAT over more than 4 years 
between 100\,MeV and 300\,GeV
is $\Gamma = 1.5 \pm 0.3$ \cite{hess2014ph}. 
Thus, the intrinsic spectral index is likely to be over-estimated by the method applied here and the results for this source can 
be regarded as conservative. The low energy threshold of CTA will guarantee a better determination of the intrinsic spectrum. 
For PKS\,1424+240 (bottom-left panel), only the coherent field of the BL Lac jet is considered. 
The rather large distance of the VHE emitting zone of $r_\mathrm{VHE} = 0.06$\,pc to the central engine 
enables a significant ALP detection above $g_{11} \gtrsim 3$ albeit the small magnetic field of $B = 0.033$\,G.
The QED vacuum polarisation effect leading to the maximal energy of eq. \eqref{eq:emax}
is only of importance for the largest values of the coupling assumed here.
For $g_{11} < 7$ it is found to be $E_\mathrm{max} = E_\mathrm{max}^\prime\delta_\mathrm{D} \gtrsim 0.9$\,TeV,
where the primed value denotes the stationary jet frame. 

\begin{figure}
\centering
\includegraphics[width = .49 \linewidth]{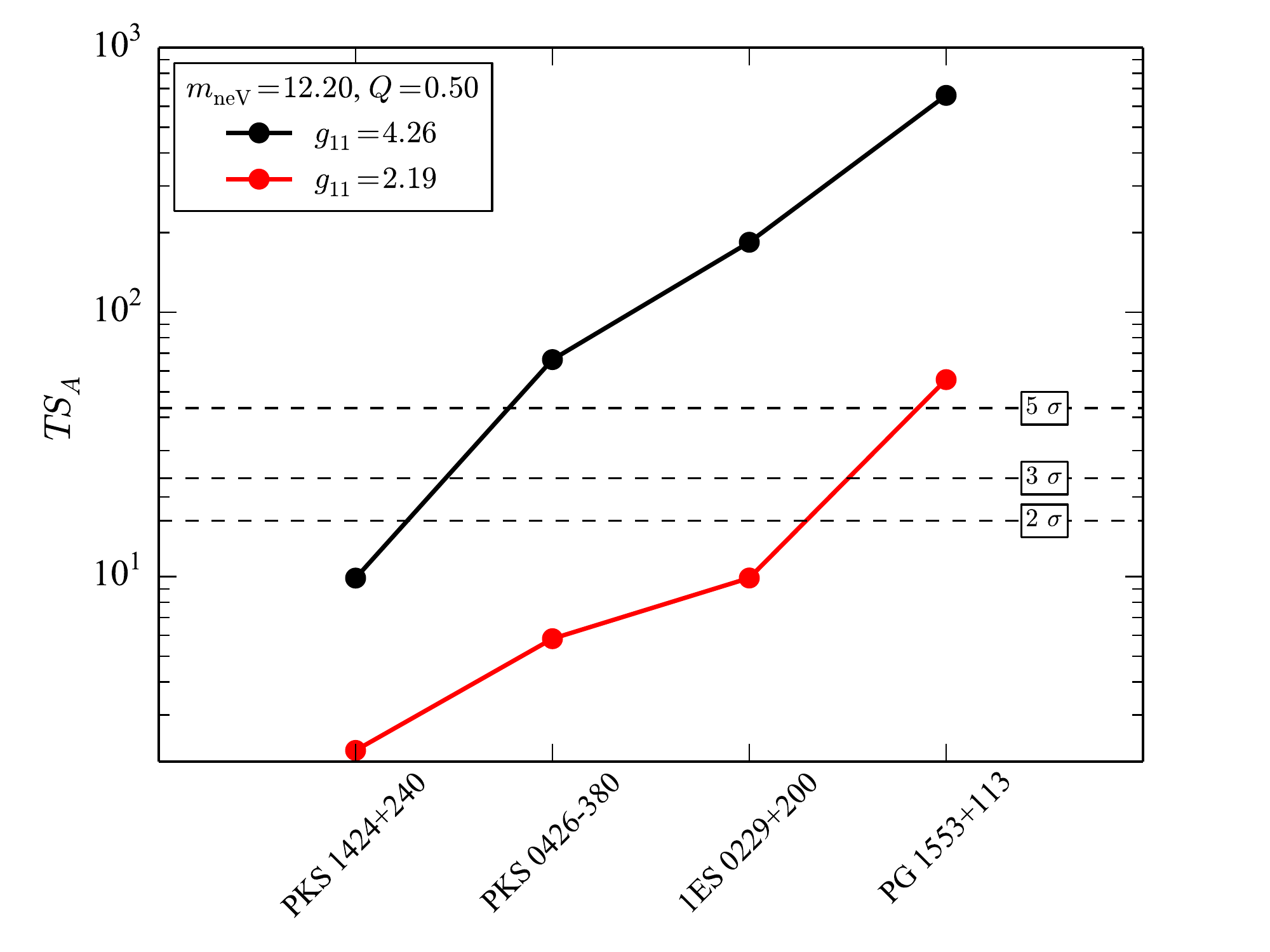}
\includegraphics[width = .49 \linewidth]{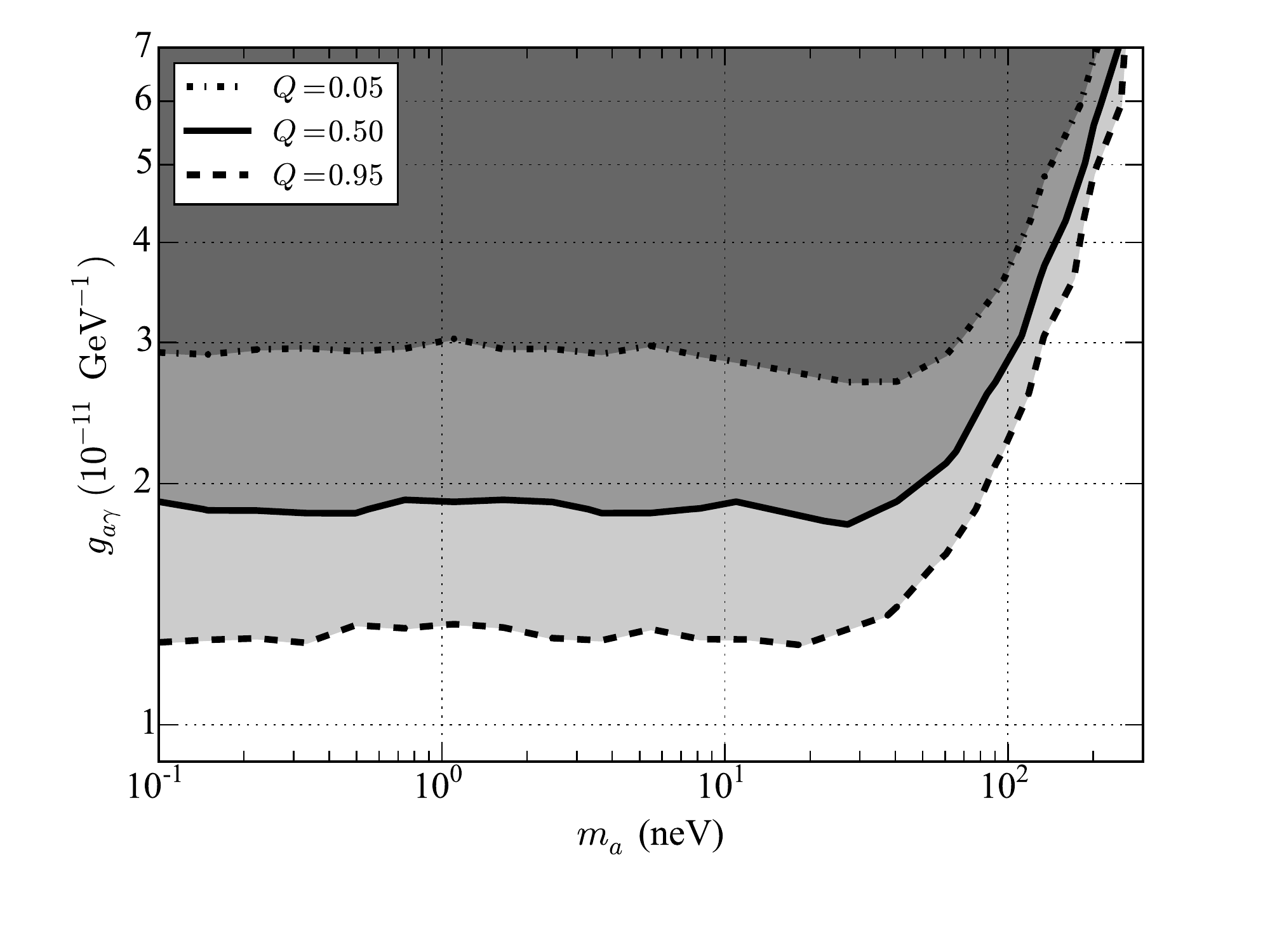}
\caption{\textit{Left:} Cumulative test statistic for two values of $g_{a\gamma}$ and fixed $m_a$ for the $Q = 0.5$ quantile.
 For each point, the $\mathit{TS}_A$ value from the given source is added.
 The dashed lines show different values of the sensitivity $\alpha_0$.
 \textit{Right:} Combined test statistic from all four blazars over the entire tested parameter range of the ALP mass and coupling.}
\label{fig:TScombined}
\end{figure}

Combining the $\mathit{TS}_A$ values from all sources increases the sensitivity to detect the ALP effect.
For one particular mass and two values of the coupling this is shown in figure \ref{fig:TScombined} (right panel), 
where the test statistics from each source 
are successively added. 
A combination of multiple sources is necessary in order to exclude source intrinsic effects for a spectral hardening. 
A boost in the $\gamma$-ray flux is unlikely to occur in many sources at exactly 
that energy where the optical depth becomes high and would correspond to an unnatural fine-tuning. 
The highest contribution to the combined test statistic comes from PG\,1553+113. 
The high flux state (cf. the spectrum in figure \ref{fig:specs}) enables the detection of the source up to very 
high $\tau$ values with high significance in 20 hours of observation time. 
The combination of the test statistic over the entire $(m_a,g_{a\gamma})$ plane for different values of $Q$ is shown in the 
left panel of figure \ref{fig:TScombined} and is clearly dominated by the results of PG\,1553+113.
The effect of the other sources only becomes visible for the $Q = 0.05$
contour line, i.e., if a $B$-field realisation is chosen for which the photon-ALP coupling is weak. 
Combining the four blazars in this case leads to an improvement of the $3\,\sigma$ detection threshold 
below $m_\mathrm{neV} \lesssim 30$. 

\begin{figure}
\centering
\includegraphics[width = .9 \linewidth]{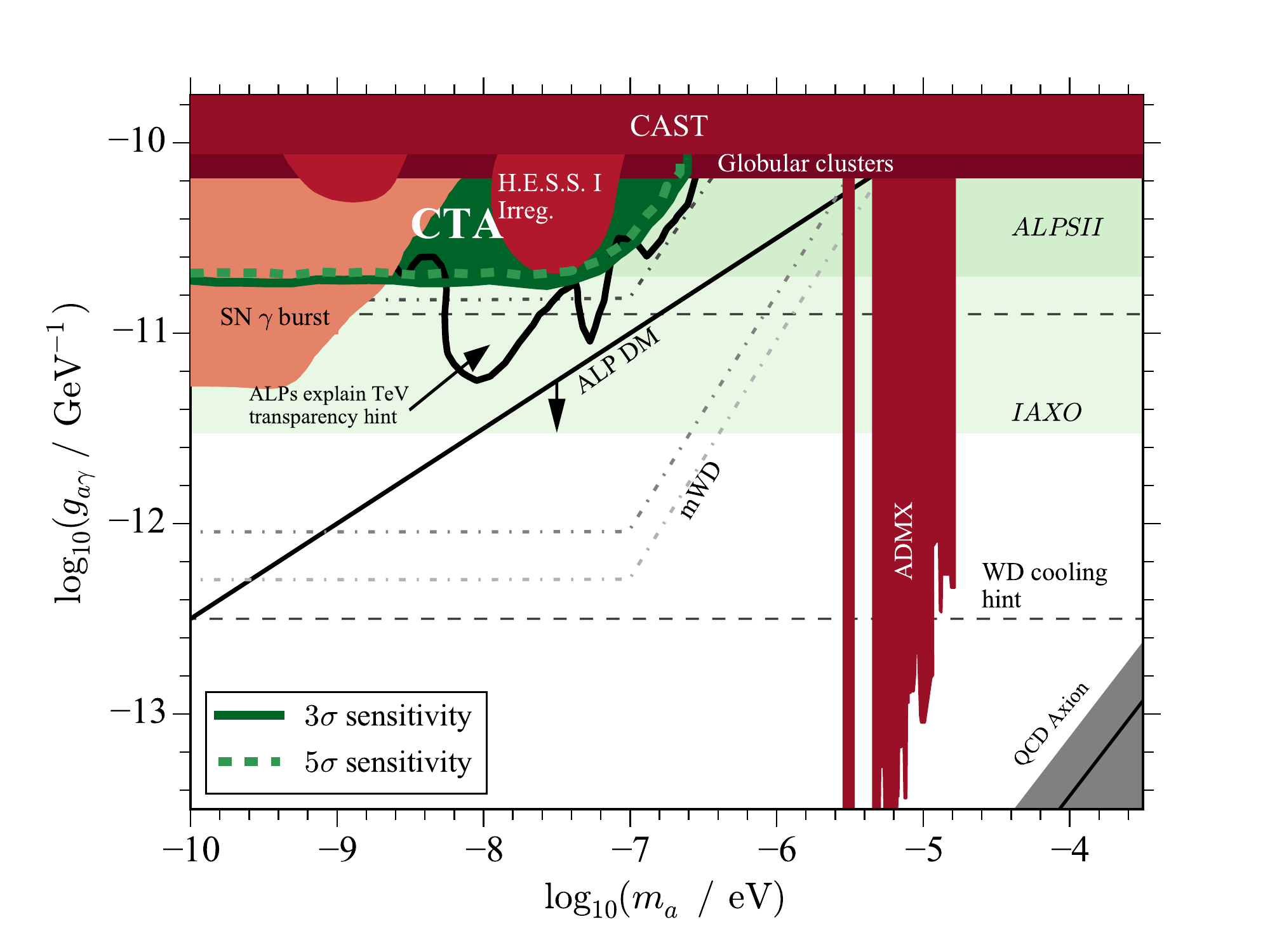}
\caption{Sensitivity at the $3$ and $5\,\sigma$ level for the $Q = 0.5$ 
quantile compared to sensitivities of other future experiments (light green regions), limits (red regions), and
theoretically preferred regions (black and grey lines and regions) in the ALP parameter space
(see the review of, e.g., ref. \cite{hewett2012} and references therein; additionally 
the results of the refs. \cite{gill2011,arias2012,hess2013:alps,ayala2014,payez2014} are added). 
}
\label{fig:limits}
\end{figure}

In figure \ref{fig:limits} we compare the 3 and 5$\,\sigma$ sensitivities
for $Q = 0.5$ with projected sensitivities of 
future experiments 
as well as with current limits and theoretically preferred regions.
Under the given model assumptions 
 CTA will be able to detect ALP induced $\gamma$-ray boosts below $m_\mathrm{neV} \lesssim 100$
and $g_{11} \gtrsim 2$. 
This is the regime where ALPs could explain hints for an opacity anomaly \cite{meyer2013}.
It should be noted that in ref. \cite{meyer2013}
lower limits on $g_{a\gamma}$ are derived and thus optimistic values for 
the magnetic fields are assumed (cluster magnetic fields are modelled with $r_\mathrm{max} = 2/3$\,Mpc, $B = 1\,\mu$G, a coherence length of $L_\mathrm{coh} = 10\,$kpc, and no decrease of the magnetic field with $r$)
together with $Q = 0.05$.
Moreover, the EBL model of ref. \cite{franceschini2008} was used 
 which is additionally upscaled by a factor of 1.3 as suggested 
by H.E.S.S. observations \cite{hess2013ebl}. 
If the same values were used in the present analysis, the ALP effect could be detected for lower values 
of the coupling. In ref. \cite{meyer2013} it is further underlined that the dips around $m_\mathrm{neV} = 10$ and $m_\mathrm{neV} = 100$ should not be interpreted as a preferred region for photon-ALP mixing since they are caused by the oscillations in $P_{\gamma\gamma}$. 
From this discussion we conclude that CTA with the observations assumed here 
will be able to probe the entire parameter space for which ALPs can explain the hints for a reduced opacity.
If a hint for an ALP is indeed observed it can be tested with the future dedicated ALP experiments 
ALPS II \cite{alpsII} and IAXO \cite{irastorza2013}.
The sensitivity derived here is compatible with the findings of ref. \cite{wouters2014}
where it was proposed to search for a correlation between AGN position and spectral hardening 
due to the conversion of ALPs into photons in the GMF. 

\section{Assessment of model assumptions} 
\label{sec:disc}

\begin{figure}
\centering
\includegraphics[width = .9 \linewidth]{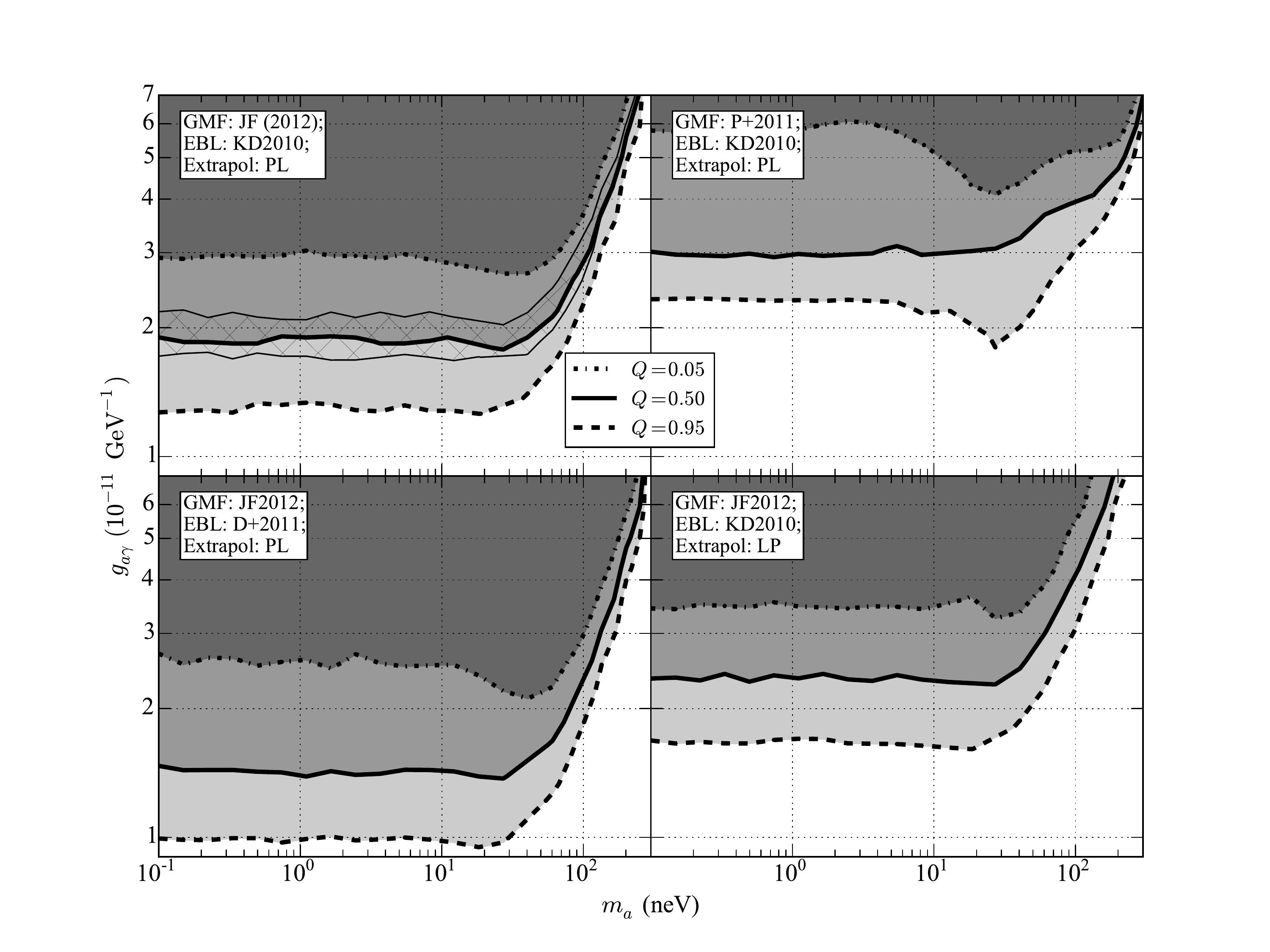}
\caption{Same as figure \ref{fig:single_src} but for the combined 
test statistic from all sources and for different assumptions 
about the EBL, GMF, and extrapolation of the intrinsic spectrum.
The hatched band for the $Q = 0.5$ quantile in the top-left panel indicates the uncertainties of  
the GMF parameters of the JF2012 model.
 See text for further details.}
\label{fig:sys-checks}
\end{figure}

 The influence of the 
different model assumptions and the uncertainty of the absolute energy scale on the final sensitivity is discussed in the following
and we provide a list of further AGN candidates to search for photon-ALP oscillations. 

\subsection{Magnetic fields at the source}
The authors of ref. \cite{meyer2014} find a strong dependence of the $\mathit{TS}_A$ 
values on the magnetic-field strength, as well as on the 
degree of turbulence of the $B$-field spectrum in the galaxy cluster and lobe scenario 
(characterised by the power-law index $q$ and the 
 minimum and maximum turbulence wave numbers $k_L$ and $k_H$ for the gaussian turbulent field, as well as the coherence length $L_\mathrm{coh}$
for the cell-like field, respectively).
The turbulence spectrum adopted here for 1ES\,0229+200 and PG\,1553+113
can be regarded as conservative, since  $q$ values above the Kolmogorov turbulence index result in a reduced  ALP production in the cluster $B$ field \cite{meyer2014}. 
Furthermore, the magnetic field in these sources is assumed to drop with increasing distance
and at 100\,kpc the field strengths read $\sim 0.12\,\mu$G and $\sim0.43\,\mu$G for PG\,1553+113 and 1ES\,0229+200, respectively.
The assumed strength of the lobe magnetic field ($1\,\mu$G) of PKS\,0426-380 are in
accordance to measured values in other sources (see, e.g., the review in ref. \cite{pudritz2012}).
The effect of ALPs should be detectable as long as the magnetic field is not 
below $\lesssim 0.5\,\mu\mathrm{G}$ or the coherence length is below $L_\mathrm{coh} \lesssim 3$\,kpc
(see ref. (2.12) and figure 5 in ref. \cite{meyer2014}).
In the BL Lac jet case, the sensitivity increases strongly with increasing distance of the VHE emitting zone to the central black hole \cite{tavecchio2014,meyer2014}.
Lacking knowledge of the exact structure of the jet and the position 
of the VHE emitting zone, the results of SSC models for the broad-band SED reflect the current best knowledge
of the $B$-field parameters of PKS\,1424+240.

\subsection{GMF model}
The errors on the best-fit parameters of the JF2012 model translate into an uncertainty of 
the reconversion probability in the GMF, $P_{a\gamma}^\mathrm{GMF}$. 
Varying all parameters but those describing the $B$ field in the galactic disk (which gives only a small contribution to the 
overall $P_{a\gamma}^\mathrm{GMF}$) one finds that the scale height of the halo field, $z_0$, 
has the strongest effect:
$P_{a\gamma}^\mathrm{GMF}$ changes almost linearly with $z_0$ for all considered sources 
with about $\pm 50\,\%$ for $z_0 \pm \sigma_{z_0}$. 
For the northern sources (PG\,1553+113 and PKS\,1424+240),
a correlation with the field strength of the halo and the second out-of-plane component (the so-called X component)
 is observed, whereas $P_{a\gamma}^\mathrm{GMF}$ anti-correlates with these parameters for the southern sources. 
As PG\,1553+113 gives the largest contribution to the overall $\mathit{TS}$ values, 
we estimate the resulting uncertainty on the sensitivity by recalculating it with all GMF parameters set to their minimum 
and maximum values within the 1\,$\sigma$ errors, respectively.
 This results in $\sim0.1\,$dex uncertainty 
on the overall sensitivity, indicated by the hatched band in the top-left panel of figure \ref{fig:sys-checks}.

Changing the GMF model to the axisymmetric (ASS) model of ref. \cite{pshirkov2011} (P+2011) leads 
to the sensitivity shown in the top-right panel of figure \ref{fig:sys-checks}. 
The $3\,\sigma$ sensitivity is reduced by $\sim60$\,\% from $g_{11} \sim 2$ to $g_{11} \sim 3$ (for $Q = 0.5$)
for the combined analysis of all sources. 
The main difference between the JF2012 and P+2011 GMF models is the absence 
of the X component in the latter one
which gives rise to an increased conversion probability for extragalactic sources at high galactic latitude.
This additional component is required in the JF2012 model 
in order to adequately describe the used data set of RM and radio synchrotron maps 
(the model results in a reduced $\chi^2 / \mathrm{d.o.f.} = 1.096 $ for 6605 data points and 21 model parameters) \cite{jansson2012}.
In terms of the fit quality, the P+2011 ASS model performs worse \cite{pshirkov2011}: A reduced $\chi^2 / \mathrm{d.o.f.} = 2.23$
(104 d.o.f.) is obtained in the northern galactic hemisphere and $\chi^2 /\mathrm{d.o.f.} = 3.74$ for 65 d.o.f. in the southern hemisphere. 
The model has especially problems explaining the RM in the direction of the galactic anti centre.
Observations in the disk are well described with $\chi^2 /\mathrm{d.o.f.} = 1.03$.
Moreover, the JF2012 model is completely divergence-free. 
Thus, we favour the JF2012 model but underline that (for $Q =0.5$) 
an ALP signal should be detectable for $g_{11} \gtrsim 3$ regardless of the GMF model assumed.
 
\subsection{EBL model}
The unknown level of the EBL photon density introduces a further uncertainty 
in our analysis. In the bottom-left panel of figure \ref{fig:sys-checks}
we repeat the analysis, this time with the EBL model of ref. \cite{dominguez2011} (D+2011)
instead of the lower limit model of KD2010. Compared to the latter one, the former model
predicts a higher photon density (at $z = 0$) at infrared wavelengths
and a lower photon density in the optical and ultraviolet.
The optical depths of the highest energy bins of the VHE spectra are (cf. section \ref{sec:sources})
$\tau = 6.72$, $\tau = 5.34$, and $\tau = 5.06$ for 1ES\,0229+200, PG\,1553+113, and PKS\,1424+240, respectively.
Not surprisingly, the stronger attenuation leads to a stronger $\gamma$-ray boost.
Consequently, the ALP effect could be detected for $g_{11} \gtrsim 1$ ($Q = 0.5$), corresponding
to an improvement of $\sim25\,\%$, re-assuring that the choice of the KD2010 model is indeed conservative.

\subsection{Extrapolation}
In the bottom-right panel of figure \ref{fig:sys-checks},
the extrapolation of the intrinsic spectrum is changed from a power law (PL) to a log parabola (LP).
The parameters of the parabola are set to $\alpha = \Gamma$ and $\beta = 0.32$. 
The de-correlation energy $E_0$ is set to the energy where $\tau = 1.8$.
The choice of $\beta$ is motivated from the curvature observed in the flaring 
spectrum of PKS\,2155-304 with H.E.S.S. \cite{hess2013:alps}. 
For comparison, the median value of $\beta$ for all sources
in the 2 year point-source catalog of the \emph{Fermi}-LAT \cite{2fgl}
 that show significant curvature is $\sim0.12$.
The reduced flux at higher energies leads to a smaller number of bins 
that enter the likelihood ratio test in eq. \eqref{eqn:loglRatio}
as they fail the $2\,\sigma$ detection threshold. 
As a result, the sensitivity is reduced by $\sim25\,$\%. 
However, such a curvature of the intrinsic spectra
would be at odds with the observed spectra in the EBL only case.

\subsection{Energy scale} 
We have tested the effect of the unknown absolute 
energy scale by downscaling all energies by 15\,\%. 
As a result, the extrapolation of the intrinsic spectrum 
reaches to optical depths below $\tau = 11$.
It should be noted that the CTA consortium aims for a lower uncertainty in the energy scale \cite{cta2011}
and that a cross correlation of currently operating IACTs with the \emph{Fermi}-LAT
using observations of the Crab Nebula results in a correction of the energy scale of about $\sim 5$\,\% \cite{meyer2010}.
 However, 
repeating the analysis yields almost the same results as before since optical depth  
remains high enough to attenuate the primary component sufficiently strong.

\subsection{Further source candidates}
It cannot be excluded that the source magnetic fields assumed here are over-estimating the true values.
For example, the magnetic field in the vicinity of the radio source NGC\,0315
which is situated in a poor galaxy group could be as low 
as $\sim 0.1\,\mu$G \cite{laing2006}, an order of magnetic below what is assumed here 
for PG\,1553+113 and 1ES\,0229+200.  
The complete degeneracy of the photon-ALP oscillation in terms of $g_{a\gamma}$ and $B$ implies 
that the coupling would have to be increased by a factor of 10 to compensate the lower magnetic field \cite{meyer2014}
which is in the regime that is ruled out by globular cluster observations \cite{ayala2014}.
Therefore, it will be important to expand the source list beyond the four blazars considered 
here in the future. 
As shown in section \ref{sec:results}, the strongest constraints come from the 
observation of a flaring state of PG\,1553+113. 
The redshift lower limit of $z \leqslant 0.4$ translates into an energy of $E_{\tau = 4} \sim 0.97$\,TeV
 (in the KD2010 model) corresponding to the energy regime around one to a few TeV
where CTA is expected to be most sensitive. 
We have searched the \emph{Fermi}-LAT catalog of
sources detected above 10\,GeV (1FHL, \cite{1fhl}) for 
additional observation candidates. 
We set initial cuts of a spectral index $\Gamma < 2.3$ and an integrated 
energy flux $> 3\times10^{-11}\,\mathrm{erg}\,\mathrm{cm}^{-2}\,\mathrm{s}^{-1}$
motivated by the index assumed here for PKS\,0426-380 and the energy flux of this source given in the 1FHL.
First, we search for
variable sources\footnote{
Variability is tested in the 1FHL with a Bayesian block analysis. 
A number of blocks larger than 1 indicates a variable source, see ref. \cite{1fhl}
for further details. 
} with redshifts between 0.2 and 0.5 which translates to an energy range $0.64\,\mathrm{TeV}\lesssim E_{\tau = 4} \lesssim 5.5\,\mathrm{TeV}$.
The results are shown in the upper part of table \ref{tab:fhl}. 
PG\,1553+113 is recovered together with 3 additional sources. It becomes clear that 
PG\,1553+113 is remarkable in the sense that from the variable sources in table \ref{tab:fhl} it has the largest energy flux 
in the energy band covered in the 1FHL and the second hardest spectrum. 
We also include one source for which no redshift is available. 
Three of the four sources are detected with IACTs and we list the maximum $\tau$ values for these
sources in the rightmost column of table \ref{tab:fhl}.
Observations of flaring states of these sources could offer the opportunity to 
search for a boost in $\gamma$-ray flux. 

Furthermore, we search the 1FHL for non-variable sources at any redshift but with the same cuts on the index and integrated energy flux as before. 
These sources are listed in the lower part of figure \ref{tab:fhl}. All of these sources are detected by IACTs. 
They have similar characteristics to 1ES\,0229+200 and PKS\,1424+240, thus
deep observations will be required in order to be able to detect an ALP induced spectral hardening.

\begin{table}
\centering
\begin{footnotesize}
\begin{tabular}{|l|ccccc|}
\hline
\multirow{2}{*}{Name} & Spectral  & Energy flux  & \multirow{2}{*}{$z$} & \multirow{2}{*}{Variability} & \multirow{2}{*}{$\tau$} \\
{} & Index & $(\times10^{-11}\mathrm{erg}\,\mathrm{cm}^{-2}\,\mathrm{s}^{-1})$ & {} & {} & {} \\
\hline
3C\,66A & $2.20\pm0.10$ & $10.34 \pm1.21$ & $\geqslant 0.3347$ & 5 & 1.76 \\
PKS\,0301-243& $2.02\pm0.17$ & $3.99 \pm0.91$ & $0.2657$ & 3 & 2.42   \\  
PG\,1553+113& $1.99\pm0.09$ & $14.67 \pm1.80$ & $\geqslant0.4$ & 2 & 3.95 \\  
PMN\,J1603-4904& $1.96\pm0.14$ & $6.13 \pm1.16$ & $?$ & 2 & -- \\ 
\hline
1ES\,0502+675& $1.63\pm0.12$ & $6.23 \pm1.15$ & $0.340$ & 1 & -- \\   
1ES\,0647+250& $1.58\pm0.18$ & $4.41 \pm1.23$ & $0.45$? & 1 & -- \\
PKS\,1424+240& $2.27\pm0.12$ & $8.39 \pm1.11$ & $\geqslant0.6035$ & 1 & 4.30 \\
\hline
\end{tabular}
\end{footnotesize}
\caption{Further potential sources to search for ALP signatures. 
The spectral index, energy flux, and variability are extracted from the 1FHL \cite{1fhl}. The variability 
is given in numbers of Bayesian blocks. If the source
is detected with IACTs, the last column gives the optical depth of the 
highest energy bin of the VHE spectrum.
1ES\,0502+675 and 1ES\,0647+250 have been detected with VERITAS \cite{benbow2011}
and MAGIC \cite{delotto2012}, respectively, but the spectra are not published upon writing.
Redshifts marked with a question mark are unknown or uncertain
(redshifts are taken from the Roma BZCAT catalog \cite{bzcat} supplemented with information from refs. \cite{danforth2010,kotilainen2011,furniss2013,shaw2013}). 
}
\label{tab:fhl}
\end{table}

\section{Summary and conclusion} 
\label{sec:concl}
We have derived the sensitivity for the proposed CTA to detect a flux enhancement
at high optical depths in $\gamma$-ray spectra from blazars caused 
by oscillations of photons into ALPs.
The joint analysis of four blazars shows that CTA will be able 
to detect an ALP signal for photon-ALP couplings $g_{a\gamma} \gtrsim 2\times10^{-11}\,\mathrm{GeV}^{-1}$ 
and ALP masses $m_a \lesssim 100$\,neV and thus will cover the 
parameter region suggested in ref. \cite{meyer2013} for which ALPs could explain the hints for a reduced opacity of the Universe.

The sensitivity depends on the assumed magnetic field scenarios and different $B$ fields close to the source including 
the fields in BL Lac jets, in the lobes of the jet, and in the environment of BL Lacs (poor galaxy clusters or galaxy groups) have 
been considered. While the dependence of the sensitivity on the parameters of the fields close to the source has been studied in ref. \cite{meyer2014},
the dependence on the galactic magnetic field, the EBL, and the assumed intrinsic blazar spectrum have been investigated here. 
The largest effect can be attributed to the chosen GMF model and the assumed intrinsic spectrum. 
For a more pessimistic GMF model the sensitivity is reduced to $g_{a\gamma} \gtrsim 3\times10^{-11}\,\mathrm{GeV}^{-1}$
while an extrapolation of the intrinsic spectrum with a logarithmic parabola instead of a power law leads 
to a significant detection above $\sim2.5\times10^{-11}\,\mathrm{GeV}^{-1}$. 
Throughout this work, an EBL model has been used that predicts a minimal attenuation at TeV energies \cite{kneiske2010}
and using instead a different model \cite{dominguez2011} increases the sensitivity by 25\,\%.
Including the uncertainty on the absolute energy scale in the analysis does not affect the results. 

The most promising targets for ALP searches are identified to be blazars in flaring states located at a redshift around $z \sim 0.4$ 
so that the optical depth is $\tau \gtrsim 4$ for an energy of 1\,TeV where CTA is expected to be most sensitive. 
From our results we find that PG\,1553+113 is a prime target, however, 
it is mandatory to include several sources in order to exclude a source intrinsic effect. 

Alternatively, CTA observations could be used to search for irregularities in $\gamma$-ray spectra. 
This method probes a narrow ALP mass range \cite{hess2013:alps} 
but has the advantage that nearby sources can be used. Prime examples are PKS\,2155+304 
or the bright radio galaxy NGC\,1275 located at the centre of the Perseus cluster. 
If evidence for photon-ALP oscillations is indeed found in CTA spectra
it will be directly testable with proposed future experiments such as ALPS II \cite{alpsII} 
and IAXO \cite{irastorza2013}.
The probed ALP mass range could be extended using $\gamma$ rays of energies beyond tens of TeV
 with observations with HAWC \cite{hawc2013} or the proposed HiSCORE
experiment \cite{tluczykont2011}.

Alternatives to photon-ALP oscillations that explain the hints for a reduced opacity 
have been put forward in the literature. 
While inhomogeneities of the EBL density do not seem to suffice as an explanation \cite{furniss2014}, 
electromagnetic cascades induced by ultra-high energy cosmic rays 
could also be produce  a spectral hardening at high optical depths \cite{essey2010,essey2011,essey2012}.
This scenario could be probed with the same observational strategy as proposed here.

\acknowledgments
We thank Daniele Montanino, Dieter Horns, and Miguel S\'anchez-Conde for providing valuable comments to the manuscript. J.C. is a Wallenberg Academy Fellow.

\bibliographystyle{JHEP}
\bibliography{meyer_ALP_transparency,galaxy_clusters,vhe_spectra}
\end{document}